\documentclass[fleqn,10pt]{article}
\usepackage[utf8]{inputenc}
\usepackage[T1]{fontenc}
\usepackage{dcolumn}
\usepackage{bm}
\usepackage{amsmath, amsthm, amsxtra, amsfonts, amssymb,amscd,mathtools,latexsym}
\usepackage{empheq} 
\usepackage{color}
\usepackage[normalem]{ulem} 
\usepackage{authblk,cite}
\usepackage[pdftex]{graphicx}

\def\eps{{\varepsilon}}
\def\R{{\mathbb R}}

\title{Synchronization and oscillation quenching in interacting metronomes on a movable platform: simple model and bifurcation analysis}

\author[1,*]{Yusuke Kato}
\author[1]{Hiroshi Kori}
\affil[1]{Department of Complexity Science and Engineering, Graduate School of Frontier Sciences, The University of Tokyo, Kashiwa, Chiba 277-8561, Japan}
\affil[*]{To whom correspondence should be addressed. Email: yuukato@g.ecc.u-tokyo.ac.jp}

\begin{document}
\maketitle
\begin{abstract}
    Various oscillatory phenomena occur in the world. Because some oscillations are related to abnormal states (e.g., particular diseases), establishing state-transition methods from an oscillatory to a resting state is important. In this study, we construct a simple metronome model and analyze the oscillation-quenching phenomenon of metronomes on a platform as an example of such state transitions. Although numerous studies were conducted on the metronome dynamics, most of them focused on the synchronization, and few studies treated the oscillation quenching because of the difficulty in analysis. 
    To facilitate the analysis, we model a metronome as a linear spring pendulum with an impulsive force (escapement mechanism) described by a fifth-order polynomial. By performing an averaging approximation, we obtain a phase diagram for the in-phase synchronization, anti-phase synchronization, and oscillation quenching. We also numerically integrate the equation of motion and confirm the agreement between the analytical and numerical results. 
    Despite the simplicity, our model successfully reproduces essential phenomena in interacting mechanical clocks, such as the bistability of in-phase and anti-phase synchrony and oscillation quenching occurring for a large mass ratio between the oscillator and the platform. We believe that our simple model will contribute to future analyses of other dynamics observed in metronomes.
\end{abstract}


Oscillatory phenomena are widely observed in nature and society. 
There are both desirable and undesirable rhythms. For example, some rhythms in the human body, such as the heartbeat and circadian clock, are essential for homeostasis, while others appear with diseases, such as abnormal oscillations of neuronal action potentials in Parkinson's disease \cite{brown2001dopamine} or epilepsy \cite{jirsa2014nature}. Thus, it is expected that stabilizing the necessary oscillations and removing the abnormal oscillations will contribute to healthy biological rhythms and the treatment of diseases.
Mathematically, an oscillatory state corresponds to the limit cycle of a dynamical system \cite{strogatz2018nonlinear}. Threfore, developing techniques to stabilize or destabilize limit cycles by external perturbations is important.

Several methods for stabilizing unstable limit cycles were established in the 1990s \cite{pyragas1992continuous,ott1990controlling}. 
On the other hand, regarding the methods to destabilize the stable limit cycles, although various state-transition methods between multiple limit cycles have been extensively studied \cite{pecora1991pseudoperiodic,yang1995trajectory,jiang1999trajectory,safonov2006noise,pisarchik2014control}, few studies have explored the general principle for state transition from stable limit cycles to stable fixed points \cite{pisarchik2000annihilation,pisarchik2001controlling,chang2020falling,chang2021falling}. 
Establishing methods for transferring the system state from a stable limit cycle to a stable fixed point is important for annihilating undesirable rhythms \cite{chang2020falling,chang2021falling}. 

In the present paper, we consider the metronome as an example of a bistable system with a stable limit cycle and stable fixed point. We focus on the oscillation-stopping phenomenon (oscillation quenching) of the metronome on a platform in which the metronome stops vibrating after oscillating for a while. Below, we explain the details of our study referring to the previous studies on metronome dynamics. 

A metronome is a mechanical device in which a needle with a weight continuously oscillates. Its characteristic structure is that the combination of a spring and gear provides torque in the same direction as the motion of the needle when the needle reaches a certain position \cite{kapitaniak2012synchronization}.
This structure, called an escapement mechanism, enables the metronome to oscillate by counteracting the damping force caused by friction. 
To investigate the various dynamical behaviors of metronomes and their mechanical analogues, e.g., pendulum clocks, numerous experimental studies have been performed \cite{pantaleone2002synchronization,bennett2002huygens,czolczynski2009clustering,czolczynski2011huygens,wu2012anti,martens2013chimera,kapitaniak2014imperfect,wu2014experimental,oliveira2015huygens,pena2016sympathy,goldsztein2022coupled}. One of the most famous dynamics of coupled metronomes is synchronization, in which the timing of the metronome oscillation is aligned when multiple metronomes are placed on a common platform. Synchronization was first discovered by Huygens in an experiment using pendulum clocks \cite{bennett2002huygens}, and many experiments have since been conducted in various settings \cite{pantaleone2002synchronization,czolczynski2009clustering,czolczynski2011huygens,wu2012anti,oliveira2015huygens,pena2016sympathy,goldsztein2022coupled}. In particular, it is known that both in-phase and anti-phase synchronizations, or only one of them, can be observed depending on the experimental situation \cite{wu2012anti}. Another unique behavior is oscillation quenching, which has been observed in metronomes on a movable platform \cite{goldsztein2022coupled} or the pendulum clocks suspended on a movable board \cite{bennett2002huygens}. However, although a lot of previous studies focused on the synchronization of metronomes, few investigated the oscillation quenching.

Numerous modeling studies have also been performed to analyze the dynamics of metronomes and pendulum clocks \cite{pantaleone2002synchronization,bennett2002huygens,moon2006coexisting,ulrichs2009synchronization,czolczynski2009clustering,czolczynski2011huygens,wu2012anti,kapitaniak2012synchronization,martens2013chimera,kapitaniak2014imperfect,wu2014experimental,oliveira2015huygens,pena2016sympathy,xin2017analysis,goldsztein2021synchronization,goldsztein2022coupled}. As summarized in Ref \cite{goldsztein2021synchronization}, several challenges exist with these studies. The first is the modeling of the escapement mechanism. To describe the escapement mechanism, it is considered appropriate to use functions that provide torque in the same direction as pendulum motion. In the previous studies, the van der Pol-type function \cite{pantaleone2002synchronization,ulrichs2009synchronization,martens2013chimera}, the piecewise linear function \cite{czolczynski2009clustering,czolczynski2011huygens,wu2014experimental,kapitaniak2014imperfect,pena2016sympathy,xin2017analysis}, Dirac's delta function \cite{goldsztein2021synchronization,goldsztein2022coupled}, and the function that instantaneously changes the angular velocity \cite{bennett2002huygens} or the kinetic energy \cite{oliveira2015huygens} of the pendulum at specific positions have been used. Another challenge is that the motion equation of a metronome generally becomes a nonlinear ordinary differential equation (ODE) that cannot be solved explicitly. To analyze the nonlinear motion equation, an averaging approximation \cite{pantaleone2002synchronization,goldsztein2021synchronization,goldsztein2022coupled} and Poincar\'{e} map \cite{bennett2002huygens,oliveira2015huygens} have been applied.

Recently, Goldsztein et al. analyzed a mathematical model of two metronomes on a movable platform and obtained a phase diagram for in-phase and anti-phase synchronization \cite{goldsztein2021synchronization}. In particular, they succeeded in explaining some of the past experiments by considering the metronome's nonlinearity caused by its pendulum structure \cite{goldsztein2021synchronization}; that is, they expanded the usual linear small-angle approximation ($\sin \theta \simeq \theta$) to include the nonlinear term ($\sin \theta \simeq \theta + c \theta^3$ with sufficiently small $c$). Seeking a better agreement with the experimental results, they also created a more realistic model by assuming Coulomb friction as the damping of the platform \cite{goldsztein2022coupled}. 
Although these studies are elaborate and sophisticated in terms of both modeling methods and analytical techniques, there are several open questions. In their first study \cite{goldsztein2021synchronization}, the analytical method (i.e., averaging approximation) was applied only when the amplitude of the metronome was larger than a certain threshold value, preventing the analysis of oscillation quenching. 
In the second study \cite{goldsztein2022coupled}, although this issue was resolved (i.e., the averaged system was valid even when the amplitude was small), a stability analysis was not performed because the averaged system contained discontinuous functions. The behavior of the equation of motion was tested only by numerical simulations, and a phase diagram was not created.

We are particularly concerned with oscillation quenching because oscillation quenching can be considered as an example of a state transition from a stable limit cycle (oscillating state) to a stable fixed point (resting state) when the oscillators receive the feedback resulting from their motion via the platform. Thus, this study aims to treat both the synchronization and oscillation quenching in a unified manner, that is, using the same mathematical model. 
One of the reasons why the analysis of oscillation quenching was difficult in the previous studies \cite{goldsztein2021synchronization,goldsztein2022coupled} is that the authors attempted to make the model realistic by modeling the escapement mechanism with a delta function and considering both the nonlinearity of the pendulum structure and the Coulomb friction acting on the platform. 
Thus, to facilitate the analysis, we model the metronome as a linear spring pendulum, neglect the damping of the platform, and simulate the escapement with several smooth functions, particularly a  polynomial of order five. Owing to these simplifications, we analytically and numerically obtain a phase diagram for both synchronization and oscillation quenching, which has not been obtained in the previous studies on metronomes.

The remainder of this paper is organized as follows. First, we consider the case of a single metronome on a movable platform. 
To analyze the equation of motion using an averaging approximation, we treat the entire system as a weakly nonlinear oscillator by assuming that both the escapement mechanism and damping are sufficiently small. 
We use several functions to represent the escapement mechanism and confirm that the averaged system reproduces the bistability and oscillation quenching of a real metronome. 
We then expand our model to the case of two identical metronomes on a movable platform, where we adopt a fifth-order polynomial as the escapement mechanism to make the analysis easier. Assuming that the mass ratio of the metronome to the platform is sufficiently small, we perform an averaging approximation and a linear stability analysis to obtain a phase diagram for the in-phase synchronization, anti-phase synchronization, and oscillation quenching. We verify the analysis by numerically integrating the equations of motion before the averaging approximation and plotting the results on the same phase diagram. 
Finally, we provide a summary and discussion. 

\section*{One metronome}
\label{sec_single_metronome}
\subsection*{Model}
\begin{figure}
\centering
\includegraphics[width=0.7\linewidth]{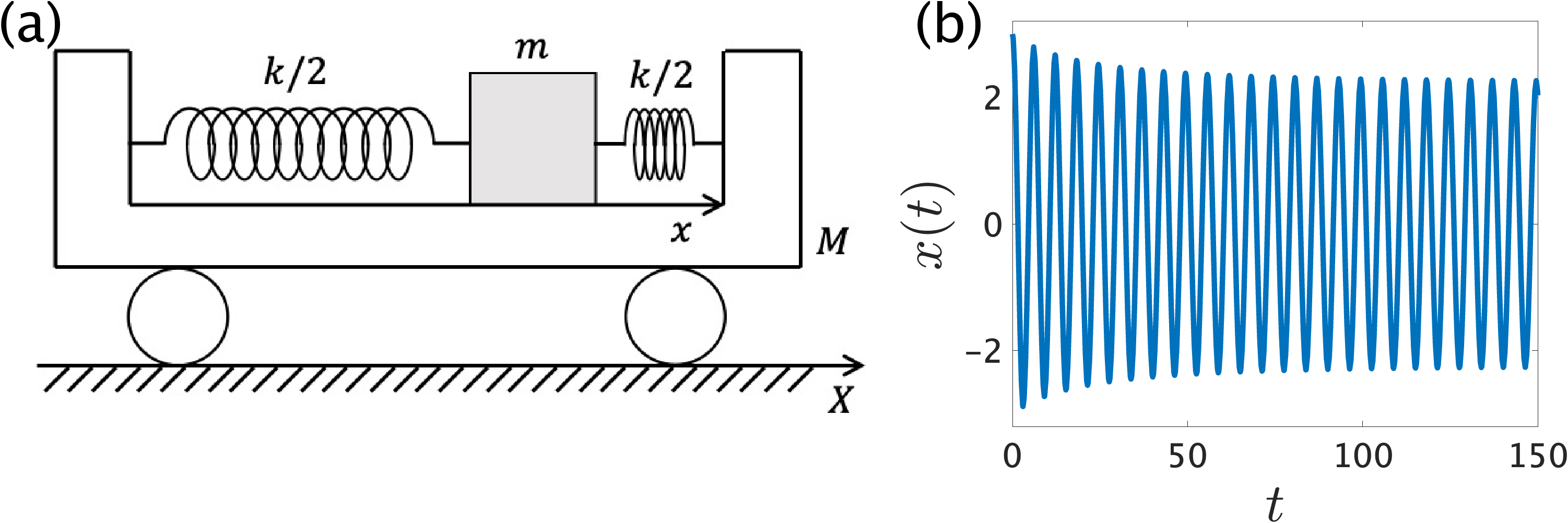}
\caption{({\bf a}) The model of one metronome on a movable platform. To facilitate the later analysis, we ignore the pendulum structure of the real metronome. ({\bf b}) Typical dynamics of the nondimensional motion equation \eqref{motion_eq_4'}. We set $g$ as Eq. \eqref{poly}, $\eps = 0.01$, $a=4$, $b=1$, $\alpha = 0.5$, and $(x(0),\dot x(0)) = (3,0)$.}
\label{model1}
\end{figure}

Figure \ref{model1} ({\bf a}) illustrates our model of one metronome on a movable platform. 
In this model, a point mass $m$ is connected by two springs with spring constant $k/2$ to a platform of mass $M$.  The platform has one degree of freedom and moves freely. The variables $X(t)$ and $x(t)$ are the positions of the platform relative to the floor and the mass relative to the platform, respectively. 

We assume the following two forces that act from the platform to the mass point: the damping force proportional to the velocity of the mass, $\dot x \coloneqq dx/dt$, and the active force describing the escapement mechanism given as a function of $x$ and $\dot x$. By neglecting the mass of the springs, the air resistance of the mass and platform, and the damping of the platform due to friction with the floor, we obtain the following motion equation: 
\begin{equation}
\label{motion_eq}
m\ddot{x} + kx + \gamma \dot x - \delta f(x,\dot{x}) + m\ddot{X} = 0, 
\end{equation}
where $\gamma$ is the damping coefficient, $\delta$ represents the magnitude of the escapement mechanism, and $f(x,\dot{x})$ is a dimensionless function that shows the nature of the escapement mechanism. 

We denote the position of the center of mass as $x_\mathrm{c}$. Then, 
\begin{equation}
x_\mathrm{c} = \frac{MX + m(X+x)}{M+m}.
\end{equation}
Since we assume that there is no external force acting on the entire system, $\frac{d^2 x_\mathrm{c}}{d t^2} = 0$ follows. Thus, 
\begin{equation}
\label{ddot_X}
\ddot{X} = -\frac{m}{M+m}\ddot{x}. 
\end{equation}
Substituting Eq. \eqref{ddot_X} into Eq. \eqref{motion_eq}, we have
\begin{equation}
\label{motion_eq_2}
\frac{Mm}{M+m}\ddot{x} + kx + \gamma \dot x - \delta f(x,\dot{x}) = 0. 
\end{equation}
We then nondimensionalize Eq. \eqref{motion_eq_2}. By introducing a small dimensionless parameter $\eps$ and the following quantities: 
\begin{equation}
    \label{nondim_quantities}
    \mu \coloneqq \frac{m}{M},\quad \omega \coloneqq \sqrt{\frac{(1 + \mu )k}{m}}, \quad \tau \coloneqq \omega t, \quad \alpha \coloneqq \frac{\omega \gamma}{\eps k}, \quad \hat x \coloneqq \frac{k x}{\delta}, \quad g \left(\hat x, \frac{d \hat x}{d \tau} \right) \coloneqq \frac{1}{\eps}f \left(\frac{\delta \hat x}{k}, \frac{\omega \delta}{k}\frac{d \hat x}{d \tau}\right). 
\end{equation}
and renaming $\tau \to t$ and $\hat x \to x$, we transform Eq. \eqref{motion_eq_2} into the following dimensionless system: 
\begin{equation}
    \label{motion_eq_4'}
    \ddot{x} + x = \eps \left(- \alpha \dot x + g(x,\dot{x}) \right). 
\end{equation}
Here, we assume $x = O(1)$, $\dot x = O(1)$, $\alpha = O(1)$ and $g(x,\dot x) = O(1)$. We show the typical dynamics of Eq. \eqref{motion_eq_4'} in Fig. \ref{model1} ({\bf b}). 

\subsection*{Analysis}
\label{sec_analysis_one}
We perform the averaging approximation to the system \eqref{motion_eq_4'}. 
We rewrite Eq. \eqref{motion_eq_4'} as 
\begin{subequations}
    \label{motion_eq_5}
    \begin{align}
        \dot x &= y, \\
        \dot y &= -x + \eps \left( - \alpha y + g(x,y) \right). 
    \end{align}
\end{subequations}
We transform the variables $x(t)$ and $y(t)$ into the new variables $r(t)$ with $r(t) \geq 0$, $\theta(t)$ , and $\phi(t)$ that satisfy
%
\begin{subequations}
    \label{change_variable2}
    \begin{align}
        x(t) &= r(t) \cos (t + \theta(t)) = r(t) \cos \phi(t), \label{change_variable2_x}\\
        y(t) &= - r(t) \sin (t + \theta(t)) = - r(t) \sin \phi(t), 
    \end{align}
\end{subequations}
where 
\begin{equation}
\label{def_phi}
    \phi(t) \coloneqq t + \theta(t).
\end{equation}
Then, Eq. \eqref{motion_eq_5} is transformed into
\begin{subequations}
    \label{polar_eq}
    \begin{align}
        \dot r &= - \eps \sin \phi \left( \alpha r \sin \phi + g(r \cos \phi, -r \sin \phi) \right), \\
        \dot \theta &= - \frac{\eps}{r} \cos \phi \left( \alpha r \sin \phi + g(r \cos \phi, -r \sin \phi) \right).
    \end{align}
\end{subequations}
See Supplementary Information for the derivation of Eq. \eqref{polar_eq}. 

Since $r(t) = O(1)$ follows from the assumptions $x(t) = O(1)$ and $y(t) = O(1)$, Eq. \eqref{polar_eq} suggests that the time-scale of $r(t)$ and $\theta(t)$ are $O(\eps^{-1})$ and thus much larger than the time-scale of metronome's oscillation period $2\pi$ (i.e., the time-scale of $x(t)$ in Eq. \eqref{change_variable2_x}). Therefore, we can safely replace $\dot r$ and $\dot \theta$ with their time average over $2 \pi$. Namely, we approximate the right-hand sides of Eq. \eqref{polar_eq} with the time average as below: 
\begin{subequations}
    \label{average_2}
    \begin{align}
        \dot r & \simeq - \frac{\eps}{2\pi} \int_0^{2\pi} dt \sin \phi \left( \alpha r \sin \phi + g(r \cos \phi, -r \sin \phi) \right) \notag \\
        & = - \frac{\eps}{2\pi} \int_0^{2\pi} d\phi \sin \phi \left( \alpha r \sin \phi + g(r \cos \phi, -r \sin \phi) \right) = -\eps \left( \frac{\alpha r}{2} + \bar g_1(r) \right),  \label{average_2_r}\\
        \dot \theta &\simeq - \frac{\eps}{2 \pi r} \int_0^{2\pi} dt \cos \phi \left( \alpha r \sin \phi + g(r \cos \phi, -r \sin \phi) \right) \notag \\
        & = - \frac{\eps}{2 \pi r} \int_0^{2\pi} d\phi \cos \phi \left( \alpha r \sin \phi + g(r \cos \phi, -r \sin \phi) \right) = - \frac{\eps}{r} \bar g_2(r), \label{average_2_theta}
    \end{align}
\end{subequations}
where
\begin{subequations}
    \label{g12_def}
    \begin{align}
    \bar g_1(r) &= \frac{1}{2 \pi} \int_0^{2\pi} d\phi \ g(r \cos \phi, -r \sin \phi) \sin \phi, \label{g12_def_1}\\
    \bar g_2(r) &= \frac{1}{2 \pi} \int_0^{2\pi} d\phi \ g(r \cos \phi, -r \sin \phi) \cos \phi. \label{g12_def_2}
    \end{align}
\end{subequations}
Note that we regard $r$ and $\theta$ as constants when we calculate the integrals in Eqs. \eqref{average_2_r} and \eqref{average_2_theta}: this is because the change of these variables during the integral interval $[0, 2\pi]$ is $O(\eps)$ and thus can be negligible in the first-order approximation. This approximation method is widely known as averaging method \cite{pikovsky2002synchronization, hale2015oscillations} (or Krylov-Bogoliubov averaging method \cite{krylov1950introduction}), and the above discussion is mathematically justified using near identity transformation \cite{Guckenheimer1983, Sanders2007}. Hereafter, we consider the approximately equal sign ($\simeq$) in Eq. \eqref{average_2} as the equal sign ($=$). 


Below, we consider concrete functions as $g(x,y)$ and obtain $\bar g_{1,2}(r)$ given by Eq. \eqref{g12_def}.

\subsubsection*{Features of escapement mechanism}
We first discuss appropriate functions to model the escapement mechanism. Considering a real metronome, it is natural to assume that $g(x,y)$ takes non-zero values if and only if $x$ and $y$ have the same sign. This is because, in the real metronome, the repulsive force works only when the pendulum position is the right from the center and moves to the right, or the pendulum position is the left from the center and moves to the left 
. Thus, we set $g(x,y)$ to match this assumption. 


Below, we describe the three cases where $g(x,y)$ is the piecewise linear function, the rational function with numerator of degree 3 and denominator of degree 4, and the 5th order polynomial, respectively. In Supplementary Information, we describe another case where $g(x,y)$ is the rational function with a linear numerator and quadratic denominator. 

\subsubsection*{Model (i)}
\label{sec_pl}
\begin{figure}
    \centering
    \includegraphics[width=.9\linewidth]{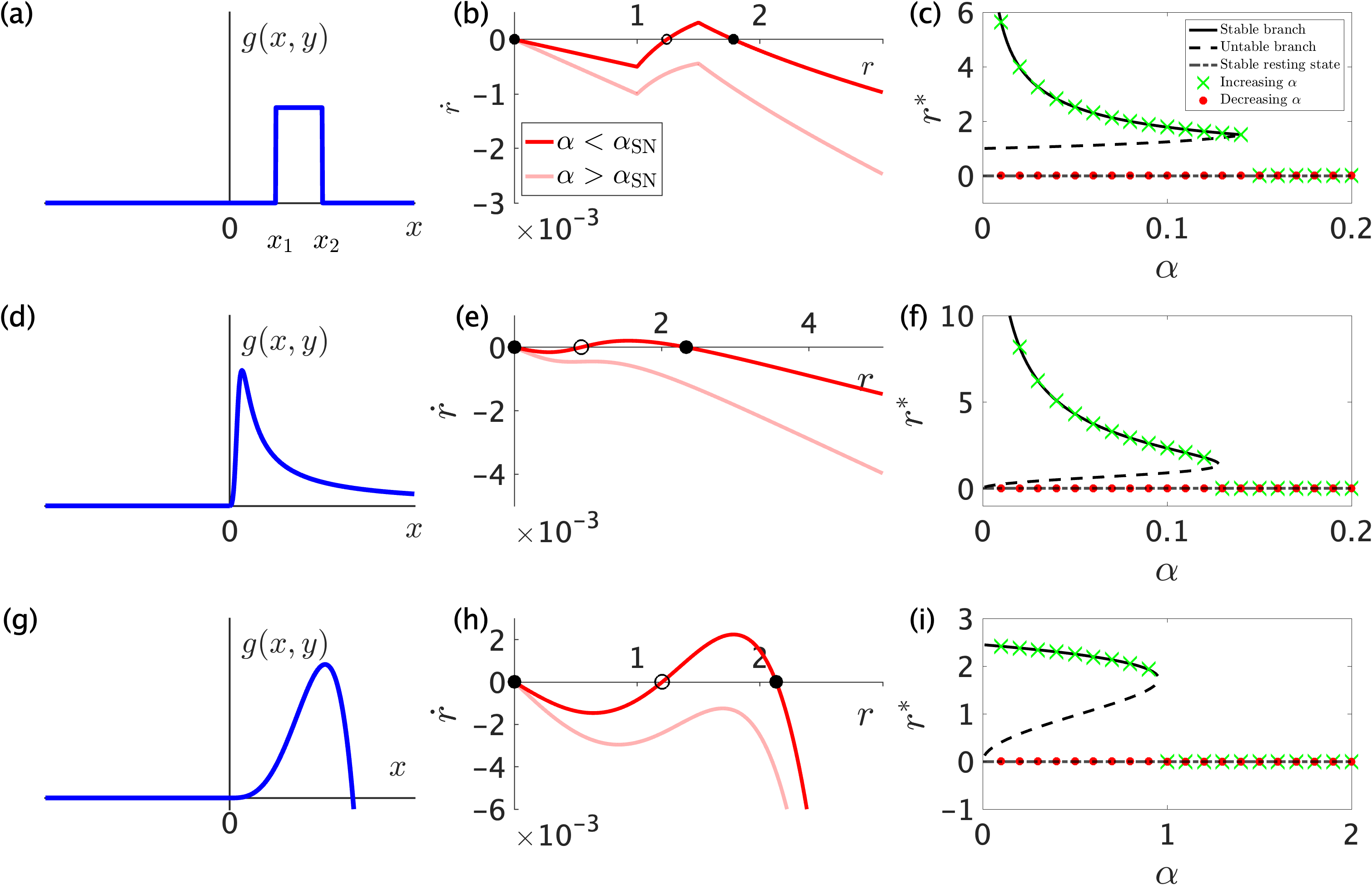}
    \caption{Left column: the shape of $g(x,y)$ with positive $y$ in Model (i) (Panel ({\bf a})), (ii) (Panel ({\bf d})), and (iii) (Panel ({\bf g})). Middle column: the flow for $r$ dynamics in Model (i) (Eq. \eqref{amp_piece}; Panel ({\bf b})), Model (ii) (Eq. \eqref{amp_sm_2}; Panel ({\bf e})), and Model (iii) (Eq. \eqref{amp_poly}; Panel ({\bf h})). The black dot and black circle in each panel represent the stable and unstable fixed points, respectively. Right column: the bifurcation diagram for $r$ obtained by both the averaging approximation and the numerical simulation of Eq. \eqref{motion_eq_4'} where $g$ is given by Model (i) (Panel ({\bf c})), Model (ii) (Panel ({\bf f})), and Model (iii) (Panel ({\bf i})). The solid, dashed, and dash-dotted lines correspond to the analytically obtained stable fixed point ($r^*=r_{\rm stable}$), the unstable fixed point ($r^* = r_{\rm unstable}$), and the trivial fixed point ($r^* = 0$), respectively. The green cross marks show the numerically obtained fixed points in the case where $\alpha$ increases, while the red dots show those in the case where $\alpha$ decreases. We first use $\alpha = \alpha_{\rm min}$ and then increase $\alpha$ by $\alpha_{\rm inc}$ until $\alpha = \alpha_{\rm max}$. Next, we decrease $\alpha$ by $\alpha_{\rm inc}$ until $\alpha = \alpha_{\rm min}$. The initial conditions for the first simulation in these three panels are the same; $x(0) = 2.0, \dot x(0) = 0$. For the following simulations, we use the fixed point of the previous simulation as the initial condition. 
    The parameters are as follows: $\eps = 0.01$ for all the panels, $x_1 = 1.0,\: x_2 = 1.5,\: \alpha = 0.1$ and $0.2$,  for panel ({\bf a}), $x_1 = 1.0,\: x_2 = 1.5,\: \alpha_{\rm min} = 0.01,\: \alpha_{\rm max} = 0.2,$ and $\alpha_{\rm inc} = 0.01$ for panel ({\bf b}), $\alpha = 0.1$ and $0.2$ for panel ({\bf c}), $\alpha_{\rm min} = 0.01,\: \alpha_{\rm max} = 0.2,$ and $\alpha_{\rm inc} = 0.01$ for panel ({\bf d}), $a=4,\: b=1,\: \alpha = 0.7$ and $1.1$ for panel({\bf e}), and $a=4,\: b=1,\: \alpha_{\rm min} = 0.1,\: \alpha_{\rm max} = 2.0,$ and $\alpha_{\rm inc} = 0.1$ for panel ({\bf f}). }
    \label{single_flow_bif}
\end{figure}

We use the following piecewise linear function, which has been previously used \cite{czolczynski2009clustering,pena2016sympathy,xin2017analysis}, as the escapement mechanism of the metronome: 
\begin{equation}
    \label{pw_li}
    g(x, y) \coloneqq 
    \begin{cases}
        1 & {\rm if}\quad x_1 < x < x_2, \ y > 0,  \\
        -1 & {\rm if}\quad -x_1 > x > -x_2, \ y < 0, \\
        0 & {\rm otherwise}, 
    \end{cases}
\end{equation}
where $x_1, x_2$ are the positive constants that satisfies $x_1 < x_2$. This is one of the simplest models of the real escapement mechanism. The shape of the function $g$ with $y>0$ is shown in Fig. \ref{single_flow_bif} ({\bf a}). 

In this case, the averaging equation \eqref{average_2} is calculated as follows: 
\begin{subequations}
    \label{amp_piece}
    \begin{empheq}[left={\dot r = \empheqlbrace}]{alignat=2}
        & -\frac{\eps \alpha}{2} r &\quad {\rm if} \quad r < x_1, \\
        & -\frac{\eps \alpha}{2} r + \frac{\eps}{\pi} \left( 1-\frac{x_1}{r} \right) &\quad {\rm if}\quad x_1 \leq r < x_2, \label{amp_piece_2} \\
        & -\frac{\eps \alpha}{2} r + \frac{\eps}{\pi} \left( \frac{x_2 - x_1}{r} \right) &\quad {\rm if} \quad r \geq x_2, \label{amp_piece_3}
    \end{empheq}  
\end{subequations}
\begin{subequations}
    \label{ph_piece}
    \begin{empheq}[left={r \dot \theta = \empheqlbrace}]{alignat=2}
        & 0 \quad {\rm if} \: r < x_1, \\
        & -\textstyle\frac{\eps}{\pi} \sqrt{1-\frac{x_1^2}{r^2}} \quad {\rm if} \: x_1 \leq r < x_2, \\
        & -\textstyle\frac{\eps}{\pi} \left(\sqrt{1-\textstyle\frac{x_1^2}{r^2}} - \sqrt{1-\textstyle\frac{x_2^2}{r^2}} \right) \quad {\rm if} \: r \geq x_2. 
    \end{empheq}
\end{subequations}
See Supplementary Information for the derivation of Eqs. \eqref{amp_piece} and \eqref{ph_piece}. 

We consider the dynamics of Eq. \eqref{amp_piece}, which is closed with respect to $r$. According to the stability analysis, described in Supplementary Information, we find the following: 
\begin{itemize}
    \item Eq. \eqref{amp_piece} has the trivial fixed point $r=0$, which is always stable regardless of the value of $\alpha$.
    \item The saddle-node bifurcation occurs at 
    \begin{equation}
        \label{pl_bif_pt}
        \alpha = \alpha_{\rm SN} \coloneqq
        \begin{cases}
            \frac{2(x_2 - x_1)}{\pi x_2^2} & {\rm if} \quad x_2 \leq 2x_1, \\
            \frac{1}{2\pi x_1} & {\rm if} \quad x_2 > 2x_1. 
        \end{cases}
    \end{equation}
    \item If $\alpha < \alpha_{\rm SN}$, Eq. \eqref{amp_piece} has the nontrivial two fixed points, one of which is stable and the other is unstable, whose values are, respectively, given by
    \begin{subequations}
    \label{r_stable_pl}
    \begin{empheq}[left={r_{\rm stable} =  \empheqlbrace}]{alignat=2}
            &\textstyle\frac{1 + \sqrt{1-2\alpha \pi x_1}}{\alpha \pi} \quad {\rm if} \quad x_2 > 2x_1, \quad \alpha \geq \textstyle\frac{2(x_2 - x_1)}{\pi x_2^2},  \label{r_stable1_pl} \\
            &\textstyle\sqrt{\frac{2(x_2 - x_1)}{\alpha \pi}} \quad {\rm otherwise}, \label{r_stable2_pl} 
    \end{empheq}
    \end{subequations}
    \begin{equation}
        \label{r_unstable_pl} 
        r_{\rm unstable} = \frac{1 - \sqrt{1-2\alpha \pi x_1}}{\alpha \pi}. 
    \end{equation}
\end{itemize}
Figure \ref{single_flow_bif} ({\bf b}) presents the typical flows described by Eq. \eqref{amp_piece}, which shows that the saddle-node bifurcation occurs as $\alpha$ changes. The bifurcation diagram for $r$ is shown in Fig. \ref{single_flow_bif} ({\bf c}). The green cross marks in Fig. \ref{single_flow_bif} ({\bf c}) show the numerically obtained equilibrium states of Eq. \eqref{motion_eq_4'} when we increase $\alpha$, whereas the red dots are those when we decrease $\alpha$. These results are in good agreement with the black lines, or the analytically obtained bifurcation diagram (i.e., Eqs. \eqref{r_stable2_pl} and \eqref{r_unstable_pl}), which validates our analytical method with averaging approximation. 

Figure \ref{single_flow_bif} ({\bf c}) indicates that for a sufficiently small alpha, our system is bistable: both the oscillatory state ($r=r_{\rm stable}$) and the resting state ($r=0$) are stable. In addition, as $\alpha$ increases, the stable limit cycle disappears by the saddle-node bifurcation and the oscillatory state transits to the resting state. Therefore, our piecewise linear model reproduces the bistability and the oscillation quenching observed in the real metronome on a movable platform.
However, in exchange for the simplicity of the model, the flow becomes non-smooth, which will actually hamper the analysis of synchronization for two metronomes. Motivated by this fact, we also consider smooth functions as $g(x,y)$ below. 

\subsubsection*{Model (ii)}
We consider
\begin{equation}
    \label{sm_2}
    g(x, y) \coloneqq 
    \begin{cases}
    \frac{x^3}{1+x^4} & {\rm if}\quad x y > 0, \\
    0 & {\rm otherwise}.
    \end{cases}
\end{equation}
The shape of the function $g$ with $y>0$ is shown in Fig. \ref{single_flow_bif} ({\bf d}). 
In this case, the averaging equations \eqref{average_2} are calculated as below: 
\begin{subequations}
    \begin{align}
        \dot r &= -\frac{\eps \alpha}{2} r + \frac{\eps}{4\pi r} \log (1+r^4), \label{amp_sm_2} \\
        r \dot \theta &= -\frac{\eps}{2r}\left( 1 - \sqrt{ \frac{1+\sqrt{1+r^4}}{2(1+r^4)}}\right). \label{ph_sm_2}
    \end{align}
\end{subequations}
See Supplementary Information for the derivation of Eqs. \eqref{amp_sm_2} and \eqref{ph_sm_2}. 

Here, we consider the dynamics of Eq. \eqref{amp_sm_2}. Obviously, the fixed point of Eq. \eqref{amp_sm_2} satisfies the following transcendental equation: 
\begin{equation}
    \label{sm_2_fix_R}
    2 \pi \alpha R = \log(1+R^2), 
\end{equation}
where $R \coloneqq r^2 \geq 0$. Since $\log(1+R^2)$ is a sigmoid function of $R$, we see that Eq. \eqref{sm_2_fix_R} has the unique solution ($R=0$) if $\alpha > \alpha_{\rm SN}$ and three solutions if $\alpha < \alpha_{\rm SN}$ (See Fig. S3 in Supplementary Information). Here, $\alpha_{\rm SN}$ is given as
\begin{equation}
    \label{sm_2_bif_pt}
    \alpha_{\rm SN} \coloneqq \frac{R^*}{\pi(1+R^{*2})}, 
\end{equation}
where $R^* > 0$ satisfies 
\begin{equation}
    \label{R*_sm_2}
    \frac{2 R^{*2}}{1 + R^{*2}} = \log(1+R^{*2}).
\end{equation}
Thus, as there exists a one-to-one relationship between $r \geq 0$ and $R$, we find the following: 
\begin{itemize}
    \item Eq. \eqref{amp_sm_2} has the trivial fixed point $r=0$, which is always stable regardless of the value of $\alpha$.
    \item The saddle-node bifurcation occurs at $\alpha = \alpha_{\rm SN}$. 
    \item If $\alpha < \alpha_{\rm SN}$, Eq. \eqref{amp_sm_2} has non-trivial two fixed points, one of which is stable ($r=r_{\rm stable}$) and the other is unstable($r=r_{\rm unstable}$), whose values are the solutions of Eq. \eqref{sm_2_fix_R} with $r_{\rm stable} > r_{\rm unstable}$. 
\end{itemize}
Figure \ref{single_flow_bif} ({\bf e}) shows the typical flows of Eq. \eqref{amp_sm_2} before and after the bifurcation point. 
The bifurcation diagram for $r$ is shown in Fig. \ref{single_flow_bif} ({\bf f}). 
The numerically obtained equilibrium states of Eq. \eqref{motion_eq_4'} (green cross marks and red dots) are in good the analytically obtained bifurcation diagram (black lines), which validates our analysis with averaging approximation. 

Figure \ref{single_flow_bif} ({\bf f}) indicates that our model with the rational function \eqref{sm_2} reproduces the bistability and the oscillation quenching observed in the real metronome on a movable platform.
However, although the flow becomes smooth in this model, we expect that the analysis of the two-oscillator system will be difficult. This is because (i) the averaged system \eqref{amp_sm_2} includes the log function and (ii) the amplitude of the stable limit cycle, which is the solution of Eq. \eqref{sm_2_fix_R}, cannot be analytically obtained. Thus, we next use the polynomial function of order 5, which partly imitates Model (i) (Eq. \eqref{pw_li}) and (ii) (Eq. \eqref{sm_2}). 

\subsubsection*{Model (iii)}
\label{sec_poly}
We consider
\begin{equation}
    \label{poly}
    g(x, y) \coloneqq 
    \begin{cases}
    a x^3 - b x^5 & {\rm if}\quad x y > 0, \\
    0 & {\rm otherwise}, 
    \end{cases}
\end{equation}
where $a , b \in \R$ are positive constants. The shape of the function $g$ with $y>0$ is shown in Fig. \ref{single_flow_bif} ({\bf g}). 
In this case, the averaging equations \eqref{average_2} are calculated as 
\begin{subequations}
    \begin{align}
        \dot r &= \frac{\eps}{12 \pi} \left( - 6 \pi \alpha r +3a r^3 - 2b r^5 \right), \label{amp_poly}\\
        r \dot \theta &= -\eps \left( \frac{3a}{16}r^3 - \frac{5b}{32}r^5 \right). \label{ph_poly}
    \end{align}
\end{subequations}
See Supplementary Information for the derivation of Eqs. \eqref{amp_poly} and \eqref{ph_poly}. 

We discuss the dynamics of Eq. \eqref{amp_poly}. Obviously, the fixed point of Eq. \eqref{amp_poly} satisfies $6 \pi \alpha r - 3a r^3 + 2b r^5 = 0$, which is solved as
\begin{equation}
    r = 0, \sqrt{\frac{3a \pm \sqrt{9 a^2 - 48 \pi b \alpha}}{4b}}. 
\end{equation}
Thus, we find the following: 
\begin{itemize}
    \item Eq. \eqref{amp_poly} has the trivial fixed point $r=0$, which is always stable regardless of the value of $\alpha$.
    \item The saddle-node bifurcation occurs at 
    \begin{equation}
        \label{poly_bif_pt}
        \alpha = \alpha_{\rm SN} \coloneqq \frac{3 a^2}{16 \pi b}. 
    \end{equation}
    \item If $\alpha < \alpha_{\rm SN}$, Eq. \eqref{amp_poly} has non-trivial two fixed points, one of which is stable and the other is unstable, whose values are, respectively, given by
    \begin{align}
        \label{r_stable_poly}
        r_{\rm stable} &= \sqrt{\frac{3a + \sqrt{9 a^2 - 48 \pi b \alpha}}{4b}}, \\
        \label{r_unstable_poly} 
        r_{\rm unstable} &= \sqrt{\frac{3a - \sqrt{9 a^2 - 48 \pi b \alpha}}{4b}}. 
    \end{align}
\end{itemize}
Figure \ref{single_flow_bif} ({\bf h}) presents the typical flows of Eq. \eqref{amp_poly} before and after the bifurcation point. 
The bifurcation diagram for $r$ is shown in Fig. \ref{single_flow_bif} ({\bf i}). 
As with the previous cases, the numerically obtained equilibrium states of Eq. \eqref{motion_eq_4'} (green cross marks and red dots) are in good the analytically obtained bifurcation diagram (black lines), which validates our analysis with averaging approximation. 

Figure \ref{single_flow_bif} ({\bf i}) indicates that model (iii) reproduces the bistability and the oscillation quenching observed in the real metronome on a movable platform. Moreover, the averaged equations \eqref{amp_poly} and \eqref{ph_poly} are expressed by the polynomial of $r$, which we expect will facilitate the analysis of the two-oscillator system. 

\section*{Two metronomes}
\label{sec_two_metronomes}
\subsection*{Model}
\begin{figure}
    \centering
    \includegraphics[width=.7\linewidth]{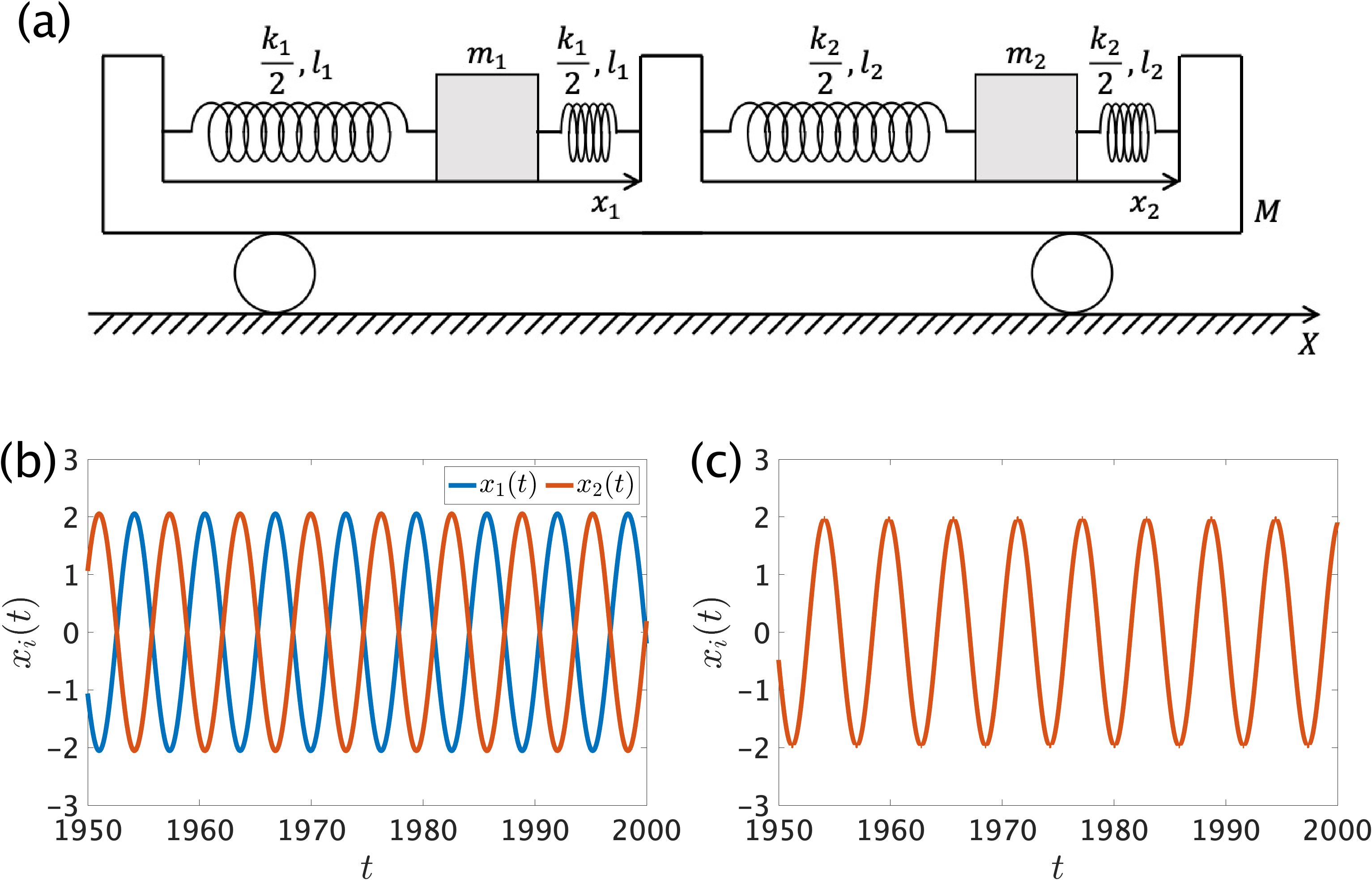}
    \caption{({\bf a}) The model of two coupled metronomes on a movable platform. ({\bf b}, {\bf c}) Typical dynamics of the nondimensional motion equation \eqref{motion_eqs_id_num}. Anti-phase and in-phase synchronization are observed in panels ({\bf b}) and ({\bf c}), respectively. We set $g$ as Eq. \eqref{poly}, $\eps = 0.01$, $a=4$, $b=1$, $\mu = 10$, and $\beta = 0.8$. Initial conditions are $(x_1(0),\dot x_1(0), x_2(0),\dot x_2(0)) = (2.8,0,-2.7,0)$ for panel ({\bf b}) and $(x_1(0),\dot x_1(0), x_2(0),\dot x_2(0)) = (2.8,0,2.7,0)$ for panel ({\bf c}).}
    \label{two_model_timeseries}
\end{figure}

We analyze the synchronization and oscillation quenching of two metronomes on a movable platform, using model (iii) as the escapement mechanism. 
The model for the two-oscillator system is shown in Fig. \ref{two_model_timeseries} ({\bf a}). 
Here, a point mass $m_i$ ($i=1$ or $2$) is connected by two springs with spring constant $k_i/2$ and natural length $l_i$ to a platform of mass $M$. The platform has one degree of freedom and moves freely. The variables $X(t)$ and $x_i(t)$ are the positions of the platform (more precisely, the position of the platform's center plate that separates the two metronomes) relative to the floor and the mass relative to the platform, respectively. The origin of $x_i$ coordinate is set to the position of the point mass $m_i$ in the equilibrium (i.e., $\dot x_1 = \dot x_2 = \dot X=0$). 

By assuming the same situation as in the single metronome model, we obtain the following motion equations:
\begin{equation}
    \label{motion_eqs}
    m_i \ddot x_i + k_i x_i + \gamma_i \dot x_i - \delta_i f_i(x_i,\dot x_i) + m_i \ddot{X} = 0, 
\end{equation}
for $i=1,2$, where $\gamma_i$ is the damping coefficient, $\delta_i$ represents the magnitude of the escapement mechanism, and $f_i(x,\dot{x})$ is a dimensionless function that shows the nature of the escapement. 

We denote the position of the center of mass as $x_\mathrm{c}$. Then, 
\begin{equation}
    x_{\rm c} = \frac{MX + m_1(X-l_1+x_1) + m_2(X + l_2 +x_2)}{M+m_1+m_2}.
\end{equation}
Since we assume that there is no external force acting on the whole system, $\frac{d^2}{d t^2}x_\mathrm{c} = 0$ holds, which implies that 
\begin{equation}
    \label{ddot_X_two}
    \ddot X = -\frac{m_1 \ddot x_1 + m_2 \ddot x_2}{M+m_1+m_2}. 
\end{equation}
Substituting Eq. \eqref{ddot_X_two} into Eq. \eqref{motion_eqs}, we have
\begin{subequations}
    \label{motion_eqs_2}
    \begin{align}
        \frac{m_1(M+m_2)}{M+m_1+m_2} \ddot x_1 + k_1 x_1 + \gamma_1 \dot x_1 - \delta_1 f_1(x_1,\dot x_1) -\frac{m_1 m_2}{M+m_1+m_2} \ddot x_2 = 0, \\
        \frac{m_2 (M+m_1)}{M+m_1+m_2} \ddot x_2 + k_2 x_2 + \gamma_2 \dot x_2 - \delta_2 f_2(x_2,\dot x_2) -\frac{m_1 m_2}{M+m_1+m_2} \ddot x_1 = 0. 
    \end{align}
\end{subequations}
We introduce a small dimensionless parameter $\eps$ and the following quantities: 
    \begin{gather}
        \omega \coloneqq \sqrt{\frac{k_1}{m_1}},\: \tau \coloneqq \omega t,\: \beta_i \coloneqq \frac{\omega \gamma_i}{\eps k_i},\: \mu_i \coloneqq \frac{m_i}{\eps M},\: 
        \hat x_i \coloneqq \frac{k_i x_i}{\delta_i},\: \kappa \coloneqq \frac{k_2}{k_1},\: \rho \coloneqq \frac{\delta_2}{\delta_1}, \:
        g_i\left(\hat x_i, \frac{d\hat x_i}{d\tau}\right) \coloneqq \frac{1}{\eps}f_i\left(\frac{\delta_i \hat x_i}{k_i}, \frac{\omega \delta_i}{k_i}\frac{d\hat x_i}{d\tau}\right). 
    \end{gather}
By renaming $\tau \to t$ and $\hat x_i \to x_i$, we transform Eq. \eqref{motion_eqs_2} into the following dimensionless system: 
\begin{subequations}
    \label{motion_eqs_3}
    \begin{align}
        \ddot x_1 + x_1 &= \eps\left(-\beta_1 \dot x_1 + g_1(x_1,\dot x_1) + \frac{\mu_1 \ddot x_1 + \frac{\rho}{\kappa} \mu_2 \ddot x_2}{1+\eps\mu_1 + \eps \mu_2}\right),  \\
        \frac{\mu_2}{\mu_1} \ddot x_2 + \kappa x_2 &= \eps\left[ - \kappa \beta_2 \dot x_2 + \kappa g_2(x_2,\dot x_2) + \frac{\mu_2(\frac{\kappa}{\rho} \mu_1 \ddot x_1  + \mu_2 \ddot x_2 )}{\mu_1(1+ \eps \mu_1 + \eps \mu_2)} \right], 
    \end{align}
\end{subequations}
where we assume that $x_i = O(1), \dot x_i = O(1), \mu_i = O(1), \beta_i = O(1), \kappa = O(1), \rho = O(1)$, and $g_i(x_i,\dot x_i) = O(1)$. 

For simplicity, we consider the case where two metronomes are identical. Namely, we set $\mu_1=\mu_2=\mu,\: \beta_1=\beta_2=\beta,\: 
\kappa = \rho = 1$, and $g_1(x,\dot x)=g_2(x,\dot x)=g(x,\dot x)$. Then, Eq. \eqref{motion_eqs_3} becomes
\begin{equation}
    \label{motion_eqs_id}
    \ddot x_i + x_i = \eps\left[-\beta \dot x_i + g(x_i,\dot x_i) + \frac{\mu(\ddot x_1 + \ddot x_2)}{1+ 2 \eps \mu} \right].
\end{equation}
Removing $\ddot x_1$ and $\ddot x_2$ terms from right-hand sides of Eq. \eqref{motion_eqs_id}, we can rewrite Eq. \eqref{motion_eqs_id} as 
\begin{equation}
    \label{motion_eqs_id_num}
        \ddot x_i = - x_i - \eps\left[\mu(x_1 + x_2) + \beta \dot x_i - g(x_i,\dot x_i) \right]  + \eps^2 \mu \left[ -\beta(\dot x_1 + \dot x_2) + g(x_1,\dot x_1) + g(x_2,\dot x_2) \right].
\end{equation}
We show the typical dynamics of Eq. \eqref{motion_eqs_id_num} in Figs. \ref{two_model_timeseries} ({\bf b}) and ({\bf c}). 
By neglecting the $O(\eps^2)$ term from the right-hand side of Eq. \eqref{motion_eqs_id_num}, we finally obtain 
\begin{subequations}
    \label{motion_eqs_id_red}
    \begin{align}
        \ddot x_1 + x_1 &= - \eps\left[\mu(x_1 + x_2) + \beta \dot x_1 - g(x_1,\dot x_1) \right],  \\
        \ddot x_2 + x_2 &= - \eps\left[\mu(x_1 + x_2) + \beta \dot x_2 - g(x_2,\dot x_2) \right]. 
    \end{align}
\end{subequations}

\subsection*{Analysis}
\subsubsection*{Averaging approximation}
We analyze Eq. \eqref{motion_eqs_id_red} with averaging approximation. We rewrite Eq. \eqref{motion_eqs_id_red} as  
\begin{subequations}
    \label{motion_eqs_5}
    \begin{align}
        \dot x_i &= y_i, \\
        \dot y_i &= -x_i - \eps \left[ \mu(x_1 + x_2) + \beta y_i - g(x_i,y_i)  \right]. 
    \end{align}
\end{subequations}
for $i=1,2$. 
We transform the variables $x_i(t)$ and $y_i(t)$ into the new variables $r_i(t)$ with $r_i(t) \geq 0$, $\theta_i(t)$ , and $\phi_i(t)$ that satisfy 
\begin{subequations}
    \label{change_variable2_two}
    \begin{align}
        x_i(t) &= r_i(t) \cos \phi_i(t), \label{change_variable2_two_x}\\
        y_i(t) &= - r_i(t) \sin \phi_i(t), 
    \end{align}
\end{subequations}
where 
\begin{equation}
\label{def_phi_two}
    \phi_i(t) \coloneqq t + \theta_i(t).
\end{equation}
Then, Eq. \eqref{motion_eqs_5} is transformed into 
\begin{subequations}
    \label{polar_eqs}
    \begin{align}
        \dot r_i &= \eps \sin \phi_i \left[ \mu(r_1 \cos \phi_1 + r_2 \cos \phi_2) - \beta r_i \sin \phi_i - g(r_i \cos \phi_i, -r_i \sin \phi_i) \right], \\
        \dot \theta_i &= \frac{\eps}{r_i} \cos \phi_i \left[ \mu(r_1 \cos \phi_1 + r_2 \cos \phi_2) - \beta r_i \sin \phi_i - g(r_i \cos \phi_i, -r_i \sin \phi_i) \right].
    \end{align}
\end{subequations}
See Supplementary Information for the derivation of Eq. \eqref{polar_eqs}. 

Since $r_i(t) = O(1)$ follows from the assumptions $x_i(t) = O(1)$ and $y_i(t) = O(1)$, Eq. \eqref{polar_eqs} suggests that the time-scale of $r_i(t)$ and $\theta_i(t)$ are $O(\eps^{-1})$ and thus much larger than the time-scale of metronome's oscillation period $2\pi$ (i.e., the time-scale of $x_i(t)$ in Eq. \eqref{change_variable2_two_x}). Therefore, we can safely replace $\dot r_i$ and $\dot \theta_i$ with with their time average over $2 \pi$, respectively, as below: 
\begin{subequations}
    \label{average_2_two}
    \begin{align}
        \dot r_1 & \simeq \frac{1}{2\pi}\int_0^{2\pi} \dot r_1 dt = \eps \biggl(\frac{\mu r_2}{2} \sin(\theta_1 - \theta_2) - \frac{\beta r_1}{2} - \bar g_1(r_1) \biggr), \\
        \dot r_2 & \simeq \frac{1}{2\pi}\int_0^{2\pi} \dot r_2 dt = \eps \biggl(\frac{\mu r_1}{2} \sin(\theta_2 - \theta_1) - \frac{\beta r_2}{2} - \bar g_1(r_2) \biggr), \\
        \dot \theta_1 & \simeq \frac{1}{2\pi}\int_0^{2\pi} \dot \theta_1 dt = \eps \biggl(\frac{\mu}{2} + \frac{\mu r_2}{2 r_1} \cos(\theta_1 - \theta_2) - \frac{1}{r_1} \bar g_2(r_1) \biggr), \\
        \dot \theta_2 & \simeq \frac{1}{2\pi}\int_0^{2\pi} \dot \theta_2 dt = \eps \biggl(\frac{\mu}{2} + \frac{\mu r_1}{2 r_2} \cos(\theta_2 - \theta_1) - \frac{1}{r_2} \bar g_2(r_2) \biggr),
    \end{align}
\end{subequations}
where $\bar g_{1,2}(\cdot)$ are given by Eq. \eqref{g12_def}. Here, based on the same arguments as in the one-oscillator system, we regard $r_i$ and $\theta_i$ as constants when we calculate the integrals in Eqs. \eqref{average_2_two}. We also use several integral formulae summarized in Supplementary Information for the derivation of Eq. \eqref{average_2_two}. Hereafter, we consider the approximately equal sign ($\simeq$) in Eq. \eqref{average_2_two} as the equal sign ($=$). 

By introducing 
\begin{equation}
    \psi \coloneqq \theta_2 - \theta_1, 
\end{equation}
Eq. \eqref{average_2_two} is rewritten as 
\begin{subequations}
    \label{average_3_two}
    \begin{align}
        \dot r_1 & = \eps \biggl(- \frac{\mu r_2}{2} \sin \psi - \frac{\beta r_1}{2} - \bar g_1(r_1) \biggr), \\
        \dot r_2 & = \eps \biggl(\frac{\mu r_1}{2} \sin \psi - \frac{\beta r_2}{2} - \bar g_1(r_2) \biggr), \\
        \dot \psi & = \eps \biggl[\frac{\mu}{2} \left(\frac{r_1}{r_2} - \frac{r_2}{r_1} \right) \cos \psi + \frac{1}{r_1} \bar g_2(r_1) - \frac{1}{r_2} \bar g_2(r_2) \biggr]. 
    \end{align}
\end{subequations}

In the analysis of the two-oscillator system, we model $g(x,y)$ with Eq. \eqref{poly} because this model reproduces the bistability and the oscillation quenching phenomenon for the one-oscillator system. Then, 
we finally obtain
\begin{subequations}
    \label{average_poly_two}
    \begin{align}
        \dot r_1 & = \frac{\eps}{12 \pi} \biggl(- 6 \pi \mu r_2 \sin \psi - 6 \pi \beta r_1 + 3a r_1^3 - 2b r_1^5 \biggr), \\
        \dot r_2 & = \frac{\eps}{12 \pi} \biggl(6 \pi \mu r_1 \sin \psi - 6 \pi \beta r_2 + 3a r_2^3 - 2b r_2^5 \biggr), \\
        \dot \psi & = \frac{\eps}{32} \biggl[16 \mu \left(\frac{r_1}{r_2} - \frac{r_2}{r_1} \right) \cos \psi + 6a(r_1^2 - r_2^2) - 5b(r_1^4 - r_2^4)\biggr]. 
    \end{align}
\end{subequations}

\subsubsection*{Analysis of synchronized states}
We perform the linear stability analysis for the averaged system \eqref{average_poly_two}. 
Eq. \eqref{average_poly_two} has two fixed points 
\begin{equation}
    \label{fixed_pt_two_in}
    (r_1, r_2, \psi) = (r^*, r^*, 0), 
\end{equation}
and
\begin{equation}
    \label{fixed_pt_two_anti}
    (r_1, r_2, \psi) = (r^*, r^*, \pi), 
\end{equation}
where
\begin{equation}
    \label{r*_two}
    r^* \coloneqq \sqrt{\frac{3a + \sqrt{9 a^2 - 48 \pi b \beta}}{4b}}. 
\end{equation}
Equations \eqref{fixed_pt_two_in} and \eqref{fixed_pt_two_anti} correspond to the in-phase and anti-phase synchronization states, respectively. Note that the condition
\begin{equation}
    \label{cond_OD}
    \beta < \beta_{\rm SN} \coloneqq \frac{3a^2}{16\pi b}
\end{equation}
is necessary for the existence of these fixed points: if $\beta \geq \beta_{\rm SN}$, these fixed points disappear with their counterparts (i.e., the fixed points $\left(\sqrt{\frac{3a - \sqrt{9 a^2 - 48 \pi b \beta}}{4b}}, \sqrt{\frac{3a - \sqrt{9 a^2 - 48 \pi b \beta}}{4b}}, 0 \right)$ and $\left(\sqrt{\frac{3a - \sqrt{9 a^2 - 48 \pi b \beta}}{4b}}, \sqrt{\frac{3a - \sqrt{9 a^2 - 48 \pi b \beta}}{4b}}, \pi \right)$ that pair with Eq. \eqref{fixed_pt_two_in} and Eq. \eqref{fixed_pt_two_anti}, respectively) by the suddle-node bifurcation. 

By performing the linear stability analysis under this parameter condition \eqref{cond_OD}, we find the following: 
\begin{enumerate}
    \item The fixed point \eqref{fixed_pt_two_anti}, corresponding to the anti-phase synchrony, is always asymptotically stable. 
    \item The fixed point \eqref{fixed_pt_two_in}, corresponding to the in-phase synchrony, is asymptotically stable if and only if 
    \begin{equation}
        \label{j3_posi}
        \mu > \mu_{\rm c} \coloneqq - \frac{15 \pi \beta}{8} + \frac{27a^2}{64b} + \frac{9a}{64b}\sqrt{9a^2 - 48 \pi b \beta}. 
    \end{equation}
\end{enumerate}
The details of the stability analysis are summarized in Supplementary Information. 

\subsubsection*{Analysis of oscillation quenching}
We first examine the stability of oscillation quenching (i.e., $x_i=\dot x_i=0$) in Eq. \eqref{motion_eqs_id_num} where $g$ is given by Eq. \eqref{poly}. In a sufficient neighborhood of oscillation quenching, Eq. \eqref{motion_eqs_id_num} can be linearized as 
\begin{equation}
    \label{motion_eqs_id_num_2}
    \ddot x_i = - x_i - \eps\left[\mu(x_1 + x_2) + \beta \dot x_i \right]  - \eps^2 \mu \beta(\dot x_1 + \dot x_2).
\end{equation}
in a first-order approximation. By introducing $z_{\rm c} \coloneqq x_1+x_2$ and $z_{\rm r} \coloneqq x_1-x_2$, Eq. \eqref{motion_eqs_id_num_2} becomes
\begin{subequations}
    \label{motion_eqs_id_num_rc}
    \begin{align}
        & \ddot z_{\rm c} + (1 + 2\eps\mu) z_{\rm c} = - \eps\beta(1+\eps^2 \mu) \dot z_{\rm c},  \\
        & \ddot z_{\rm r} + z_{\rm r} = - \eps\beta \dot z_{\rm r}.   
    \end{align}
\end{subequations}
Since Eq. \eqref{motion_eqs_id_num_rc} is the equation of damped oscillation, we see that the fixed points $z_{\rm r} = \dot z_{\rm r} =0$ and $z_{\rm c} = \dot z_{\rm c} =0$, which correspond to oscillation quenching, are asymptotically stable. 
Therefore, we expect that the averaged system \eqref{average_poly_two} converges to the oscillation quenching state in the parameter region
\begin{equation}
    \label{cond_OD_2}
    \beta > \beta_{\rm SN}, 
\end{equation}
where the fixed points that correspond to the synchronous states (i.e., Eqs. \eqref{fixed_pt_two_in} and \eqref{fixed_pt_two_anti}) disappear. 

In fact, the parameter space in which oscillation quenching occurs is wider than the inequality \eqref{cond_OD_2} if the two metronomes move in in-phase synchronization. 
Here, we consider the case when two metronomes have nearly identical movements. By assuming $x_1 = x_2 = x$, the original motion equation 
\eqref{motion_eqs_id} is transformed into
\begin{equation}
    \label{motion_eq_OD_in_phase}
    \frac{1}{1+2\eps \mu}\ddot x + x = \eps (-\beta \dot x + g(x,\dot x)). 
\end{equation}
By setting $\tau \coloneqq t \sqrt{1+2\eps \mu}$ and renaming $\tau \to t$, we obtain
\begin{equation}
    \label{motion_eq_OD_in_phase_2}
    \ddot x + x = \eps (-\beta \sqrt{1+2\eps \mu} \dot x + g(x,\dot x)). 
\end{equation}
Note that $g(x,\dot x) = g(x,\dot x \sqrt{1+2\eps \mu})$ since $g$ is given by Eq. \eqref{poly}.
In this case, we can remove the assumption that $\mu = O(1)$. In other words, as long as $\eps \mu = O(1)$, we can treat Eq. \eqref{motion_eq_OD_in_phase_2} as a weakly nonlinear oscillator and thus perform averaging approximation. 

Referring to Eqs. \eqref{amp_poly} and \eqref{ph_poly}, we see that the averaging approximation of Eq. \eqref{motion_eq_OD_in_phase_2} yields
\begin{subequations}
    \begin{align}
        \dot r &= \frac{\eps}{12 \pi} \left( - 6 \pi \beta \sqrt{1+2\eps \mu} r +3a r^3 - 2b r^5 \right), \label{amp_poly_in_phase}\\
        r \dot \theta &= -\eps \left( \frac{3a}{16}r^3 - \frac{5b}{32}r^5 \right), \label{ph_poly_in_phase}
    \end{align}
\end{subequations}
where $r$ and $\theta$ are given by Eqs. \eqref{change_variable2} and \eqref{def_phi}. Considering the dynamics of Eq. \eqref{amp_poly_in_phase}, we find that the saddle-node bifurcation occurs when
\begin{equation}
    \beta = \beta_{\rm SN\_in} \coloneqq \frac{3 a^2}{16\pi b \sqrt{1+2\eps \mu}}, 
\end{equation}
which implies that oscillation quenching is observed when 
\begin{equation}
    \label{cond_OD_in_phase}
    \beta > \beta_{\rm SN\_in}. 
\end{equation}
Note that the condition \eqref{cond_OD_in_phase} is looser than the condition \eqref{cond_OD_2}. Namely, there exist parameter regions in which the oscillation quenching occurs when two metronomes move in in-phase synchronization while anti-phase synchronization is stable, which has been observed in the previous experiments with pendulum clocks \cite{bennett2002huygens}.

\begin{figure}
    \centering
    \includegraphics[width=.9\linewidth]{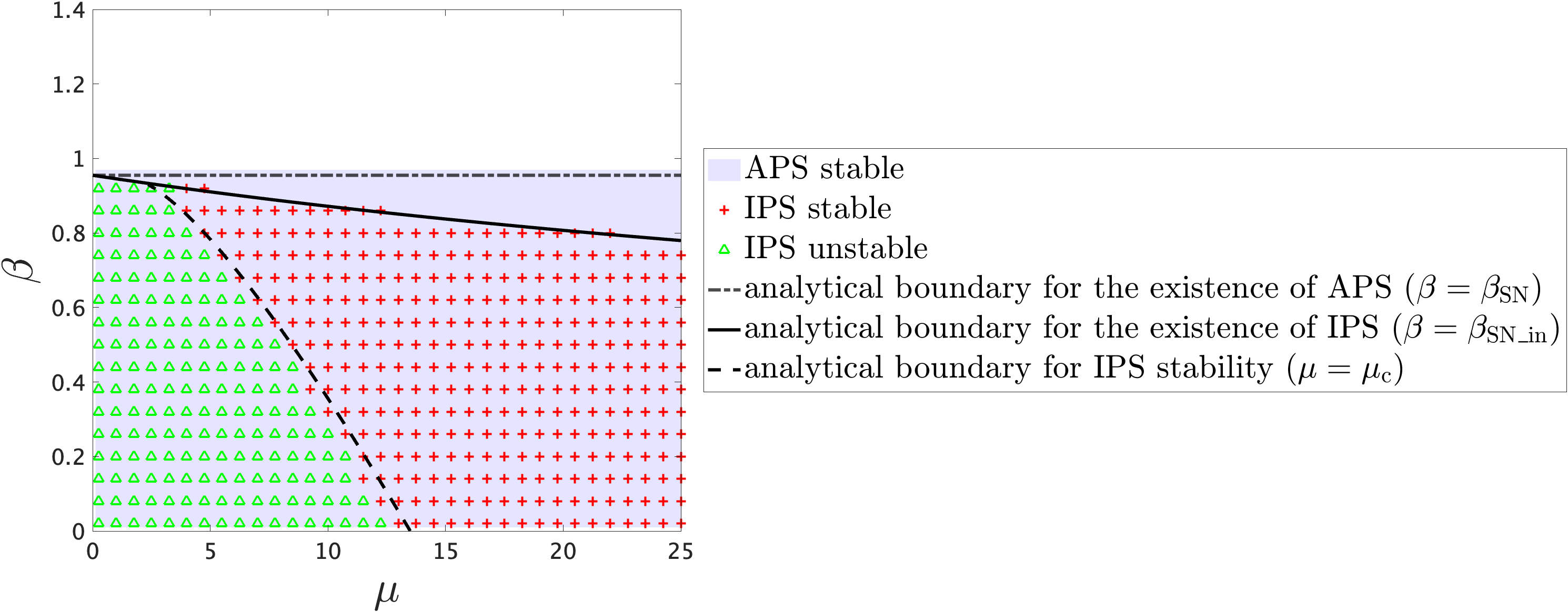}
    \caption{The phase diagram of the system \eqref{motion_eqs_id}. The black lines represent the analytical boundary obtained by an averaging approximation. The dash-dotted and solid lines are given by $\beta = \beta_{\rm SN}$ and $\beta = \beta_{\rm SN\_in}$, respectively. Namely, the dash-dotted line shows the boundary of whether oscillation quenching occurs when the two metronomes move in anti-phase synchronization, whereas the solid line is the boundary of whether oscillation quenching occurs when the two metronomes move in in-phase synchronization. The stability of in-phase synchronization switches at the dashed line, which is given by $\mu = \mu_{\rm c}$. Note that oscillation quenching is asymptotically stable in all of the parameter space. 
    The colored regions, including the white area, show the phase diagram obtained by the numerical simulation of Eq. \eqref{motion_eqs_id_num} where $g$ is given by Eq. \eqref{poly}. 
    In the white area, oscillation quenching occurs if we use the initial condition that is close to either the in-phase or anti-phase synchronization. 
    The anti-phase synchronization (APS) is stable in the blue area.
    The in-phase synchronization (IPS) is unstable in the green triangle mark area and stable in the red cross mark area. 
    Note that, in the area painted in blue only, oscillation quenching occurs if we use the initial condition that is either exactly the in-phase synchronization or close to the in-phase synchronization, whereas the anti-phase synchronization is stable. We fix $a=4, b=1, \eps = 0.01$ in the simulation. }
    \label{fig_phaseplane_tol9}
\end{figure}

Based on the above analyses, we depict the phase diagram, which is shown as the black lines in Fig. \ref{fig_phaseplane_tol9}. The dash-dotted line shows the line $\beta = \beta_{\rm SN}$, which is the boundary of whether oscillation quenching occurs when the two metronomes move in anti-phase synchronization. In the same way, the solid line shows the line $\beta = \beta_{\rm SN\_in}(\mu)$, which is the boundary of whether oscillation quenching occurs when the two metronomes move in in-phase synchronization. The dashed line is given by $\mu = \mu_{\rm c}(\beta)$ in inequality \eqref{j3_posi}. In other words, the stability of in-phase synchronization switches at the dashed line. 

The solid and dash-dotted lines in Fig. \ref{fig_phaseplane_tol9} correspond to the saddle-node bifurcation. Since the Jacobian matrix at the fixed point \eqref{fixed_pt_two_in} has a zero eigenvalue on the dashed line (see Supplementary Information), this line corresponds to either of the saddle-node, transcritical, or pitchfork bifurcation \cite{Wiggins2003}. According to the symmetry of Eq. \eqref{average_poly_two} (i.e., Eq. \eqref{average_poly_two} is invariant if we change $r_1 \to r_2,\: r_2 \to r_1$ and $\psi \to -\psi$) and the fact that this fixed point \eqref{fixed_pt_two_in} does not disappear after the bifurcation, the saddle-node and transcritical bifurcations are unlikely. Since the stable fixed point that satisfies $r_1 \neq r_2$ emerges near the bifurcation point (Fig. \ref{inphase_instable_2} ({\bf c})), we consider that the dashed line in Fig. \ref{fig_phaseplane_tol9} corresponds to the supercritical pitchfork bifurcation. 

\subsection*{Numerical simulation}
\label{sec_numerical_sim}
We verify the averaging approximation by numerically integrating Eq. \eqref{motion_eqs_id_num} and investigate the steady state for different values of $\beta$ and $\mu$, which are summarized in the colored regions in Fig. \ref{fig_phaseplane_tol9}. In the numerical simulation, we fix $a=4, b=1, \eps = 0.01$ and use the following 4 initial conditions: 
\begin{enumerate}
    \item Near in-phase synchronization 
    ($x_1(0) = 5.01,\: x_2(0) = 5,\: \dot x_{1,2} (0) = 0$),
    \item In-phase synchronization 
    ($x_1(0) = x_2(0) = 5,\: \dot x_{1,2} (0) = 0$),
    \item Near anti-phase synchronization 
    ($x_1(0) = 5.01,\: x_2(0) = -5,\: \dot x_{1,2} (0) = 0$),
    \item Anti-phase synchronization 
    ($x_1(0) = 5,\: x_2(0) = -5,\: \dot x_{1,2} (0) = 0$).
\end{enumerate}

The white area of Fig. \ref{fig_phaseplane_tol9} represents the parameter region in which oscillation quenching occurs for any of the four initial conditions. In the blue area, both the initial conditions 3 (near anti-phase synchronization) and 4 (anti-phase synchronization) converge to the anti-phase synchronization, which implies that the anti-phase synchronization is asymptotically stable in this area. In the green triangle mark area, the initial condition 2 (in-phase synchronization) converges the in-phase synchronization, whereas the initial condition 1 (near in-phase synchronization) does not, which means that the in-phase synchronization is the unstable equilibrium state of the system. In the red cross mark area, both the initial conditions 1 and 2 converge to the in-phase synchronization, meaning that the in-phase synchronization is asymptotically stable. Note that oscillation quenching occurs for the initial conditions 1 and 2 in the area painted in blue only. Namely, the in-phase synchronized steady state does not exist in this area. 

In Fig. \ref{fig_phaseplane_tol9}, we see that the analytically and numerically obtained phase diagrams are in good agreement, which confirms the validity of the averaging approximation in this study. However, the boundary for the stability of in-phase synchronization shows a slight difference between analysis (the dashed line) and numerical simulation (the boundary between the green triangle marks and the red cross marks). 
To investigate the cause of this gap, we numerically obtained $\mu_{\rm c}$, the value of $\mu$ at which the stability of the in-phase synchronization switches, while changing $\eps$ and fixing $\beta$. The results are shown in Fig. S4 in Supplementary Information, which shows that $\mu_{\rm c}$ approaches the analytically obtained value (i.e., $\mu_{\rm c}(\beta)$ in inequality \eqref{j3_posi}) as $\eps$ decreases. Thus, we conclude that the magnitude of the small parameter $\eps$ causes the gap between the boundaries of the stability of in-phase synchronization in Fig. \ref{fig_phaseplane_tol9}. 

\begin{figure}
    \centering
    \includegraphics[width=\linewidth]{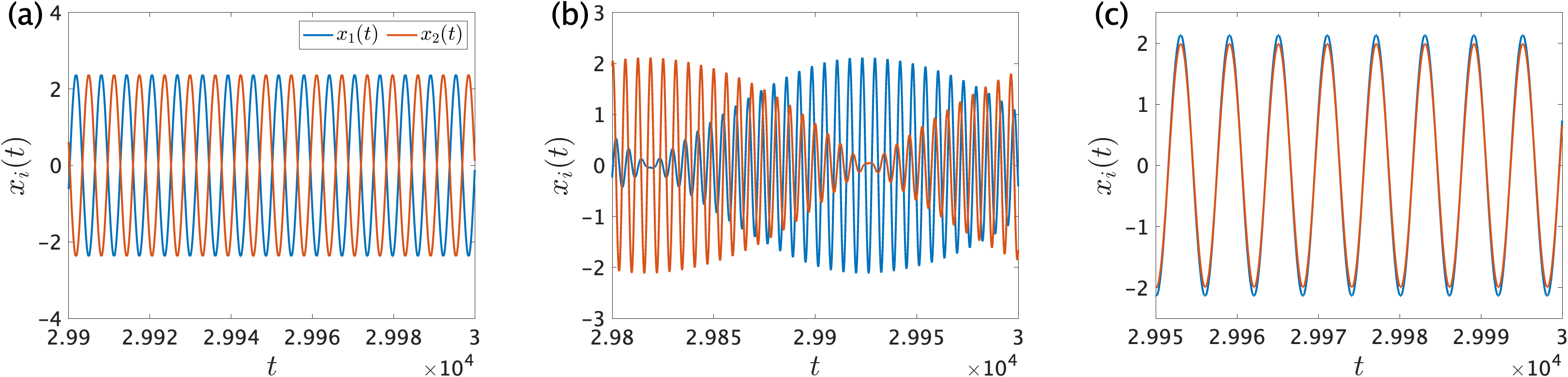}
    \caption{The equilibrium states of Eq. \eqref{motion_eqs_id_num} when we use an initial condition that is near the unstable in-phase synchronization. We numerically integrate Eq. \eqref{motion_eqs_id_num} and plot the dynamics after a sufficiently long time. We fix $a=4,\: b=1,\: \eps = 0.01,\: x_1(0) = 5.01,\: x_2(0) = 5,\: \dot x_{1,2} (0) = 0$ in this figure. ({\bf a}) When $\mu = 3, \beta = 0.2$, the system converge to the anti-phase synchronization. ({\bf b}) When $\mu = 3, \beta = 0.8$, the amplitude of each metronome alternately repeats increase and decrease. ({\bf c}) When $\mu = 5, \beta = 0.74$, that is near the bifurcation point (i.e., near the dashed line in Fig. \ref{fig_phaseplane_tol9}), near in-phase movement with slightly different amplitudes is observed. }
    \label{inphase_instable_2}
\end{figure}

In the parameter region where in-phase synchronization is unstable (i.e., the green triangle mark area in Fig. \ref{fig_phaseplane_tol9}), several equilibrium states are numerically observed when we use the initial condition 1. These results are shown in Fig. \ref{inphase_instable_2}, in which we select several sets of parameters $\beta, \mu$ and plot the time series of $x_i(t)$ in Eq. \eqref{motion_eqs_id_num} after a sufficiently long time. In our simulation results, either of the anti-phase synchronization (Fig. \ref{inphase_instable_2} ({\bf a})), the alternating increase and decrease of amplitudes which is similar to the beat phenomena in the acoustic wave (Fig. \ref{inphase_instable_2} ({\bf b})), or the near in-phase movement with slightly different amplitudes (Fig. \ref{inphase_instable_2} ({\bf c})) are observed. In particular, the dynamics in Figs. \ref{inphase_instable_2} ({\bf b,c}) suggest that there exist limit cycles and other fixed points than Eqs. \eqref{fixed_pt_two_in} and \eqref{fixed_pt_two_anti} in the averaged system \eqref{average_poly_two}, which is an important research theme for future analysis. 

\section*{Discussion}
\label{sec_discussion}
In this study, we investigate the dynamics of a metronome, which is an example of a bistable system with a stable limit cycle and a stable fixed point. In particular, we focus on the oscillation quenching phenomenon observed in metronomes on a movable platform. 
First, we construct a mathematical model for a single-oscillator system, ignoring the nonlinearity of the pendulum structure and the damping of the platform. 
By performing the averaging approximation, we consider several functions to represent the escapement mechanism and find that these functions are appropriate for simulating the real metronome's bistability and oscillation quenching. We conclude that the fifth-order polynomial is the most suitable for both the ease of analysis and reproducibility of the real metronome. 
Subsequently, we expand our model to include a two-oscillator system. By assuming that the mass ratio of the metronome to the platform, the escapement mechanism, and the damping force are sufficiently small, we perform an averaging approximation and reduce the system to a three-dimensional dynamical system of the amplitudes $r_{1,2}$ and phase difference $\psi$. By performing a stability analysis on the averaged system, we obtain the phase diagram for the in-phase synchronization, anti-phase synchronization, and oscillation quenching. 
We also numerically integrate the original motion equations and confirm agreement between the numerically and analytically obtained phase diagrams.

Our mathematical model appears to oversimplify real metronomes on a movable platform. 
However, the motion equations of our model reproduce behaviors similar to those observed experimentally, such as in-phase synchronization, anti-phase synchronization, and oscillation quenching. Thus, we believe that our simplification does not impair the essence of reality. 

The analytical method used in this study (i.e., averaging approximation) has been used in previous studies such as those by Pantaleone \cite{pantaleone2002synchronization} and Goldsztein et al. \cite{goldsztein2021synchronization,goldsztein2022coupled} However, oscillation quenching was difficult to analyze in the models adopted in these studies. This is because the former study \cite{pantaleone2002synchronization} used the van der Pol-type function to describe the escapement mechanism, which made the resting state unstable, and the latter studies \cite{goldsztein2021synchronization,goldsztein2022coupled} used the delta function, which restricted the detailed analysis of oscillation quenching. We consider several other functions (i.e., Model (i)-(iii)) for the escapement mechanism and show that the averaged system becomes bistable. In particular, for a fifth-order polynomial (Model (iii)), we can explicitly derive the amplitude of the stable limit cycle (Eq. \eqref{r_stable_poly}).
This is expected to ease the analysis. Thus, we adopt Model (iii) to describe the escapement of the two-oscillator system and successfully obtain the analytical phase diagram including the area where oscillation quenching occurs.

When considering a real metronome, an unimodal function, wherein the value is zero at the origin and infinity, seems appropriate to model the escapement mechanism. Therefore, Model (ii) is consistent with reality, although the analysis becomes difficult. To facilitate the analysis, we choose to model the escapement mechanism with a 5th-order polynomial (Model (iii)). This model captures the dynamics of the real escapement mechanism near the origin, i.e., $ax^3 - bx^5$ with $x>0$ equals zero at $x=0$ and reaches its maximum at $x=\sqrt{a/b}$. However, it diverges negatively at infinity, which means that a large unrealistic negative restoring force acts at a location far from the origin. In general, an infinite degree is required to expand a function that converges to zero at infinity into a power series. Thus, it is natural that we cannot correctly reproduce the behavior of the real escapement mechanism at infinity when we model it using a polynomial of a finite degree.

Our study shows that oscillation quenching occurs by a saddle-node bifurcation when the mass ratio $\mu$ increases. This is evident from Eq. \eqref{nondim_quantities} in the case of one metronome (note that the bifurcation parameter $\alpha$ increases as $\mu$ increases) and Fig. \ref{fig_phaseplane_tol9} in the case of two metronomes. As $\mu$ corresponds to the magnitude of the feedback that the metronome receives from the platform, our findings indicate that the oscillation can be stopped by the feedback resulting from the motion of the oscillator.

The occurrence of oscillation quenching by a saddle-node bifurcation was already reported in a previous study \cite{goldsztein2022coupled} for the case of a metronome on a fixed platform. However, our study reveals that the same bifurcation is observed even for a metronome on a movable platform. Further, we analytically find that oscillation quenching occurs when $\mu$ increases if two metronomes on a movable platform move almost in-phase synchronization, which has not been shown in previous studies \cite{goldsztein2021synchronization,goldsztein2022coupled}. In previous experiments with two pendulum clocks suspended from a common plate \cite{bennett2002huygens}, it was observed that, when two clocks start from the initial condition close to in-phase synchronization, oscillation quenching occurs as the mass ratio between the pendulum clock and the entire system increases. This observation agrees with our results, which suggests that the simplification in our modeling and the analysis with the averaging approximation are valid. 

Regarding the state-transition method from a limit cycle to a stable fixed point, our study suggests that an oscillation can be stopped by increasing the feedback resulting from its motion (i.e., the feedback from the platform). In our model, the magnitude of the feedback depends on the mass ratio $\mu$; thus, abnormal oscillations can be controlled by increasing $\mu$ instead of directly intervening in the metronome. Figure \ref{fig_phaseplane_tol9} also indicates that any perturbation that switches from anti-phase to in-phase synchronization can evoke oscillation quenching in certain parameter regions (that is, the area highlighted in blue only in Fig. \ref{fig_phaseplane_tol9}). 

There are several open questions in this study. (1) In the analysis of the averaged system \eqref{average_poly_two}, we do not clarify the existence and stability of equilibrium states other than the two fixed points $(r^*,r^*,0)$ and $(r^*,r^*,\pi)$. 
However, the steady state in which the amplitude of each metronome varies periodically is shown in Fig. \ref{inphase_instable_2} ({\bf b}), suggesting that the system \eqref{average_poly_two} has stable limit cycles. Furthermore, previous experiments revealed a phenomenon called ``metronome suppression'' \cite{goldsztein2022coupled}, in which one metronome oscillates with a larger amplitude than the other. This observation can be analyzed by performing a stability analysis of a fixed point that satisfies $r_1 \neq r_2$, as shown numerically in Fig. \ref{inphase_instable_2} ({\bf c}). Thus, it is important to investigate the stability of the steady states in the system \eqref{average_poly_two} other than synchronization and oscillation quenching. 
(2) When modeling a metronome, we neglect the nonlinearity caused by the pendulum structure. There are two reasons for this simplification: 1) the analysis becomes easier, and 2) such nonlinearity does not change the dynamics of amplitude $r$ after the averaging approximation. 
The second reason is explained as follows. In the model that considers the weak nonlinearity of the pendulum structure \cite{goldsztein2021synchronization,goldsztein2022coupled}, term $\eps \sigma x^3$ with a real constant $\sigma$ is added to the motion equation as a result of the Taylor expansion of $\sin x$ to the third-order term. Namely, Eq. \eqref{motion_eq_5} would be
\begin{subequations}
    \label{motion_eq_nlin}
    \begin{align}
        \dot x &= y, \\
        \dot y &= -x + \eps \left( - \alpha y + \sigma x^3 + g(x,y) \right), 
    \end{align}
\end{subequations}
which implies that 
\begin{subequations}
    \label{polar_eq_nlin}
    \begin{align}
        \dot r &= - \eps \sin \phi \left( \alpha r \sin \phi + \sigma r^3 \cos^3 \phi + g(r \cos \phi, -r \sin \phi) \right), \\
        \dot \theta &= - \frac{\eps}{r} \cos \phi \left( \alpha r \sin \phi + \sigma r^3 \cos^3 \phi + g(r \cos \phi, -r \sin \phi) \right).
    \end{align}
\end{subequations}
Then, the averaging approximation in Eq. \eqref{polar_eq_nlin} yields
\begin{subequations}
    \label{average_2_nlin}
    \begin{align}
        \dot r &= -\eps \left( \frac{\alpha r}{2} + \bar g_1(r) \right),  \label{average_2_r_nlin}\\
        \dot \theta &= - \frac{\eps}{r} \left( \frac{3}{8} \sigma r^3 + \bar g_2(r) \right). \label{average_2_theta_nlin}
    \end{align}
\end{subequations}
Comparing Eq. \eqref{average_2} with Eq. \eqref{average_2_nlin}, we see that the dynamics of $r$ are the same, whereas the dynamics of $\theta$ change. Because this study mainly addresses the state transition from a limit cycle to a stable fixed point and thus focuses on $r$ dynamics, we neglect the nonlinearity of the metronome, which does not change the dynamics of $r$. However, as $\theta$ dynamics are related to the phase of the oscillator, the results of the stability analysis for the two-oscillator system would change if we adopt Eq. \eqref{motion_eq_nlin} as the metronome's motion equation. In a future study, we plan to perform the same analysis for a model that considers the nonlinearity of the pendulum structure. 

We investigate the dynamics of metronomes on a movable platform using an averaging approximation and numerical simulation. To facilitate the analysis, we ignore the nonlinearity caused by the pendulum structure of the metronome and model the escapement mechanism using a fifth-order polynomial. Finally, we obtain a phase diagram and find that oscillation quenching occurs when the mass ratio between the metronome and platform increases, which agrees with previous experimental results \cite{bennett2002huygens}. We believe that our simple model will contribute to future analyses of other dynamics, such as clustering \cite{czolczynski2009clustering}, the chimera states \cite{martens2013chimera,kapitaniak2014imperfect}, and chaotic dynamics \cite{ulrichs2009synchronization}.

\section*{Methods}
All of the numerical simulations in this article were performed with MATLAB ODE45 solver. Both the absolute and relative tolerances are set to 1e$-$9 ($10^{-9}$). 

\bibliography{metronome}
\bibliographystyle{unsrt}

\section*{Acknowledgements}

This study was supported by JSPS KAKENHI (No. JP23KJ0756) to Y.K. and JSPS KAKENHI (No. JP21K12056) to H.K.

\section*{Author contributions statement}

Y.K. and H.K. conceived this project. Y.K. and H.K. designed the research methods. Y.K. performed the analysis and the numerical simulations. Y.K. drafted the manuscript under the mentorship of H.K. 

\section*{Additional information}

\textbf{Competing interests}: The authors declare no competing interests. 

\end{document}


\maketitle
\section{Derivation of Eq. \eqref{polar_eq}} 
\label{derive_me}
We introduce the complex variable $A(t)$ that satisfies 
\begin{subequations}
    \label{change_variable}
    \begin{align}
        x(t) &= \frac{1}{2} (A(t) e^{it} + {\rm c.c.} ), \\
        y(t) &= \frac{1}{2} (iA(t) e^{it} + {\rm c.c.} ).   
    \end{align}
\end{subequations}
Note that ${\rm c.c.}$ denotes the complex conjugate. 
Then, it follows from Eq. \eqref{change_variable} that 
\begin{equation}
    A(t) = (x - iy)e^{-it}, 
\end{equation}
which implies that 
\begin{equation}
\label{A_sub}
    \dot A(t) = \left[ (\dot x -y) - i (\dot y + x) \right] e^{-it}. 
\end{equation}
By substituting Eq. \eqref{motion_eq_5} into Eq. \eqref{A_sub}, we obtain 
\begin{equation}
    \label{amp_eq}
    \dot A = i e^{-it} \eps \left( \alpha y - g(x,y) \right). 
\end{equation}
We express $A(t)$ in a polar coordinate as below: 
\begin{equation}
    \label{polar}
    A(t) = r(t) e^{i \theta(t)},
\end{equation}
where $r(t), \theta(t)$ are real variables with $r(t) \geq 0$. Then, Eq. \eqref{change_variable2} follows from Eqs. \eqref{change_variable} and \eqref{polar}. 
By substituting Eqs. \eqref{polar} and \eqref{change_variable2} into Eq. \eqref{amp_eq}, we have
\begin{equation}
    \dot r + i r \dot \theta = - i e^{-i \phi} \eps 
    \left( \alpha r \sin \phi + g(r \cos \phi, -r \sin \phi) \right), 
\end{equation}
and thus we obtain Eq. \eqref{polar_eq}. 

\section{Several integral formulae with trigonometric functions}
In the derivation of Eqs. \eqref{average_2} and \eqref{average_2_two}, we use 
\begin{subequations}
    \label{trigono_avr}
    \begin{align}
        & \int_0^{2\pi} d\phi \sin^2 \phi = \int_0^{2\pi} d\phi \cos^2 \phi = \pi, \\
        & \int_0^{2\pi} d\phi \sin \phi \cos \phi = 0, 
    \end{align}
\end{subequations}
and 
\begin{align}
    \int_0^{2\pi} \sin \phi_i \cos \phi_j dt &= \int_0^{2\pi} \frac{\sin(\phi_i + \phi_j) + \sin(\phi_i - \phi_j)}{2}dt = \pi \sin(\theta_i - \theta_j), \\
    \int_0^{2\pi} \sin \phi_i \sin \phi_j dt &= \int_0^{2\pi} \frac{\cos(\phi_i - \phi_j) - \cos(\phi_i + \phi_j)}{2}dt = \pi \cos(\theta_i - \theta_j), \\
    \int_0^{2\pi} \cos \phi_i \cos \phi_j dt &= \int_0^{2\pi} \frac{\cos(\phi_i + \phi_j) + \cos(\phi_i - \phi_j)}{2}dt = \pi \cos(\theta_i - \theta_j), 
\end{align}
for $i \neq j$. Remind that $\phi_i$ is given by Eq. \eqref{def_phi_two} and $\theta_i$ is regarded as constant during the integral. 

\section{Calculation of $\bar g_{1,2}(r)$ in the case where $g$ is given as Eq. \eqref{pw_li}}
\label{pwl}
By substituting Eq. \eqref{change_variable2} into Eq. \eqref{pw_li}, we have
\begin{equation}
    \label{pw_li_app}
    g(r \cos \phi, - r \sin \phi) 
    \coloneqq 
    \begin{cases}
        1 & {\rm if}\quad x_1 < r \cos \phi < x_2, \ r \sin \phi < 0,  \\
        -1 & {\rm if}\quad -x_1 > r \cos \phi > -x_2, \ r \sin \phi > 0, \\
        0 & {\rm otherwise}. 
    \end{cases}
\end{equation}
We calculate the integrals in Eq. \eqref{g12_def} by dividing them into the following three cases: (i) $0 \leq r < x_1$, (ii) $r > x_2$, and (iii) $x_1 < r < x_2$. 

\vspace{0.5\baselineskip}
\noindent (i) In the case where $0 \leq r < x_1$, 
\begin{equation*}
    g(r \cos \phi, - r \sin \phi) = 0
\end{equation*}
holds over the integral interval (i.e., $\phi \in [0,2\pi]$), which implies that 
\begin{equation}
    \label{pwl_g12_case1}
    \bar g_1(r) = \bar g_2(r) = 0. 
\end{equation}

\vspace{0.5\baselineskip}
\noindent (ii) In the case where $r > x_2$, Eq. \eqref{pw_li_app} can be written as
\begin{equation}
    \label{pw_li_app_2}
    g(r \cos \phi, - r \sin \phi) 
    \coloneqq 
    \begin{cases}
        1 & {\rm if}\quad 2n \pi - \phi_1 < \phi < 2n \pi - \phi_2,  \\
        -1 & {\rm if}\quad (2n+1)\pi - \phi_1 < \phi < (2n+1)\pi - \phi_2, \\
        0 & {\rm otherwise}, 
    \end{cases}
\end{equation}
where 
\begin{equation}
    \phi_{i} \coloneqq \mathrm{Arccos} \frac{x_i}{r}
\end{equation}
and $n$ is an arbitrary integer. Then, 
\begin{subequations}
    \label{pwl_g12_case2}
    \begin{align}
        \bar g_1(r) &= \frac{1}{2 \pi} \left( \int_{2\pi - \phi_1}^{2\pi - \phi_2} \sin\phi - \int_{\pi - \phi_1}^{\pi - \phi_2} \sin\phi \right) \notag \\
        &= \frac{x_1 - x_2}{\pi r}, \\
        \bar g_2(r) &= \frac{1}{2 \pi} \left( \int_{2\pi - \phi_1}^{2\pi - \phi_2} \cos\phi - \int_{\pi - \phi_1}^{\pi - \phi_2} \cos\phi \right) \notag \\
        &=\frac{1}{\pi} \left( \sqrt{1-\frac{x_1^2}{r^2}} - \sqrt{1-\frac{x_2^2}{r^2}} \right). 
    \end{align}
\end{subequations}

\vspace{0.5\baselineskip}
\noindent (iii) The case where $r > x_2$ corresponds to the special case of the previous case: we only have to substitute 0 into  $\phi_2$ (i.e.,  to substitute $r$ into $x_2$) in Eqs. \eqref{pw_li_app_2} and \eqref{pwl_g12_case2}. Thus, 
\begin{subequations}
    \label{pwl_g12_case3}
    \begin{align}
        \bar g_1(r) &= 2 \left(\frac{x_1}{r} - 1 \right), \\
        \bar g_2(r) &= 2 \sqrt{1-\frac{x_1^2}{r^2}}. 
    \end{align}
\end{subequations}
\section{Stablity analysis of Eq. \eqref{amp_piece}}
\label{append_pl_stability}
Obviously, Eq. \eqref{amp_piece} has a trivial fixed point $r=0$, which is stable because $\dot r < 0$ holds in a sufficient neighborhood of $r=0$. By differentiating the right-hand-side of Eq. \eqref{amp_piece} by $r$, we obtain
\begin{align}
    \label{amp_piece_dif} 
    \frac{d \dot r}{dr} &= 
    \begin{cases}
         -\frac{\eps \alpha}{2} & {\rm if} \quad r < x_1, \\
         - \eps \left(\frac{\alpha}{2} - \frac{x_1}{\pi r^2} \right) &{\rm if}\quad x_1 \leq r < x_2, \\
        -\frac{\eps \alpha}{2} - \frac{\eps (x_2 - x_1)}{\pi r^2} &{\rm if} \quad r \geq x_2.
    \end{cases} 
\end{align}
It is easily seen that $\frac{d \dot r}{dr} < 0$ if $r < x_1$ or $r \geq x_2$, which implies that $\dot r$ is monotonically decreasing function of $r$ in these intervals. We also find that $\frac{d \dot r}{dr} = 0$ holds if and only if $r = \sqrt{\frac{2x_1}{\pi \alpha}}$ and $x_1 < \sqrt{\frac{2x_1}{\pi \alpha}} < x_2$. Below, we consider the increase or decrease of $\dot r$ in the interval $x_1 \leq r < x_2$ by dividing the cases whether $\sqrt{\frac{2x_1}{\pi \alpha}}$ belongs to this interval. 

\vspace{0.5\baselineskip} 
\noindent (i) We first consider the case where $\sqrt{\frac{2x_1}{\pi \alpha}} < x_1$, which is equivalent to 
\begin{equation}
    \alpha > \frac{2}{\pi x_1}.
\end{equation}
In this case, $\frac{d \dot r}{dr} < 0$ when $x_1 \leq r < x_2$, which implies that $\dot r$ is a monotonically decreasing function of $r$ for the whole interval $r \in [0,\infty)$. Thus, $r=0$ is the only fixed point of Eq. \eqref{amp_piece}. 

\vspace{0.5\baselineskip} 
\noindent (ii) Next, we consider the case where $x_1 \leq \sqrt{\frac{2x_1}{\pi \alpha}} < x_2$, which is equivalent to 
\begin{equation}
    \label{alpha_case2}
    \frac{2 x_1}{\pi x_2^2} < \alpha \leq \frac{2}{\pi x_1}.
\end{equation}
In this case, 
\begin{align}
    \max_{x_1 \leq r < x_2} \dot r &= \left. \dot r \right|_{r = \sqrt{\frac{2x_1}{\pi \alpha}}} \notag \\ 
    &= \frac{\eps}{\pi}\left( 1 - \sqrt{2\pi \alpha x_1} \right). \label{dotr_max_x1x2}
\end{align}
We further divide the case by the sign of the right-hand-side of Eq. \eqref{dotr_max_x1x2}. 

\vspace{0.5\baselineskip} 
\noindent (ii-a) If $\alpha \geq \frac{1}{2\pi x_1}$, then $\displaystyle\max_{x_1 \leq r < x_2} \dot r \leq 0$ follows, which implies that $r=0$ is the only fixed point of Eq. \eqref{amp_piece}. Recall that $\dot r$ is a monotonically decreasing function for the intervals $r < x_1$ and $r \geq x_2$. 

\vspace{0.5\baselineskip} 
\noindent (ii-b) If $\alpha < \frac{1}{2\pi x_1}$, then $\displaystyle\max_{x_1 \leq r < x_2} \dot r > 0$ follows. According to the inequality \eqref{alpha_case2}, this case occurs only when 
\begin{equation}
    \frac{1}{2\pi x_1} > \frac{2 x_1}{\pi x_2^2} 
    \iff x_2 > 2x_1. \label{x2_2x1}
\end{equation}
In this case, new fixed points (one is stable and the other is unstable) of Eq. \eqref{amp_piece} appear by the saddle-node bifurcation, whose coordinates are given by
\begin{subequations}
    \label{r_stable_pl_derive}
    \begin{empheq}[left={r_{\rm stable} = \empheqlbrace}]{alignat=2}
            &\textstyle\frac{1 + \sqrt{1-2\alpha \pi x_1}}{\alpha \pi} &\quad {\rm if} \quad \alpha \geq \textstyle\frac{2(x_2 - x_1)}{\pi x_2^2},  \label{r_stable1_pl_derive} \\
            &\textstyle\sqrt{\frac{2(x_2 - x_1)}{\alpha \pi}} &\quad {\rm if} \quad \alpha < \textstyle\frac{2(x_2 - x_1)}{\pi x_2^2}, \label{r_stable2_pl_derive} 
    \end{empheq}
\end{subequations}
and
\begin{equation}
    \label{r_unstable_pl_derive} 
    r_{\rm unstable} = \frac{1 - \sqrt{1-2\alpha \pi x_1}}{\alpha \pi}, 
\end{equation}
respectively. Note that the right-hand-side of Eqs. \eqref{r_stable1_pl_derive} and \eqref{r_unstable_pl_derive} are the roots of the right-hand-side of Eq. \eqref{amp_piece_2}, and the right-hand-side of Eq. \eqref{r_stable2_pl_derive} is the positive root of the right-hand-side of Eq. \eqref{amp_piece_3}. We also mention that Eq. \eqref{r_stable_pl_derive} is divided by the sign of 
\begin{equation}
    \left. \dot r \right|_{r = x_2} = -\frac{\eps x_2}{2} \left[ \alpha - \frac{2(x_2 - x_1)}{\pi x_2^2} \right].
\end{equation}

\vspace{0.5\baselineskip} 
\noindent (iii) Finally, we consider the case where $ \sqrt{\frac{2x_1}{\pi \alpha}} \geq x_2$, which is equivalent to 
\begin{equation}
    \label{alpha_case3}
    \alpha \leq \frac{2 x_1}{\pi x_2^2}.
\end{equation}
In this case, $\dot r$ is a monotonically increasing function for the interval $x_1 \leq r < x_2$. Then, we see that Eq. \eqref{amp_piece} has the nontrivial two fixed points, one of which is stable and the other is unstable, if and only if 
\begin{equation}
    \left. \dot r \right|_{r = x_2} > 0 
    \iff \alpha < \frac{2(x_2 - x_1)}{\pi x_2^2}. \label{dotr_x2_posi}
\end{equation}
We also find that the coordinates of the stable and unstable fixed points are given as Eq. \eqref{r_stable2_pl_derive} and Eq. \eqref{r_unstable_pl_derive}, respectively. 
Note that if the inequality \eqref{x2_2x1} holds, then the inequality \eqref{alpha_case3} satisfies the inequality \eqref{dotr_x2_posi} (i.e., $\frac{2 x_1}{\pi x_2^2} < \frac{2(x_2 - x_1)}{\pi x_2^2}$ follows if $x_2 > 2 x_1$). 

Based on the above discussion, we clarify the dynamics of Eq. \eqref{amp_piece}, which are summarized in the main article. 

\section{The case where $g$ is given by rational function with a linear numerator and quadratic denominator}
\label{sm_1_sec}
Here, we describe another case where $g$ is given by the following smooth rational function to model the escapement mechanism: 
\begin{equation}
    \label{sm_1}
    g(x, y) \coloneqq 
    \begin{cases}
    \frac{x}{1+x^2} & {\rm if}\quad x y > 0, \\
    0 & {\rm otherwise}.
    \end{cases}
\end{equation}
By substituting Eq. \eqref{change_variable2} into Eq. \eqref{sm_1}, we obtain
\begin{align}
    \label{sm_1_rt}
    g(r \cos \phi, -r \sin \phi) = 
    \begin{cases}
    \displaystyle\frac{r \cos \phi}{1 + r^2 \cos ^2 \phi} & 
    \begin{aligned}
        & {\rm if} \quad (2n + \textstyle\frac{1}{2}) \pi < \phi < (2n + 1) \pi \\
        & \quad {\rm or} \quad (2n - \textstyle\frac{1}{2}) \pi < \phi < 2n \pi, 
    \end{aligned} 
    \\
    \\
    0 & {\rm otherwise},
    \end{cases}
\end{align}
with arbitrary integer $n$. 
Then, it follows from Eqs. \eqref{g12_def} and \eqref{sm_1_rt} that 
\begin{align}
    \bar g_1(r) &= \frac{1}{2 \pi} \left( \int_{\frac{\pi}{2}}^{\pi} \frac{r \cos \phi \sin \phi d\phi }{1 + r^2 \cos ^2 \phi} + \int_{\frac{3}{2}\pi}^{2\pi} \frac{r \cos \phi \sin \phi d\phi}{1 + r^2 \cos ^2 \phi} \right) \notag \\
    &= \frac{-1}{4 \pi r} \left\{ \left[ \log(1+r^2 \cos ^2 \phi) \right]_{\frac{\pi}{2}}^{\pi}  + \left[ \log(1+r^2 \cos ^2 \phi) \right]_{\frac{3}{2}\pi}^{2 \pi} \right\} \notag \\
    &= \frac{- \log(1+r^2)}{2 \pi r}, \label{sm1_g1}
\end{align}
and
\begin{align}
    \bar g_2(r) & = \frac{1}{2 \pi} \left(\int_{\frac{\pi}{2}}^{\pi} \frac{r \cos^2 \phi d\phi }{1 + r^2 \cos ^2 \phi} + \int_{\frac{3}{2}\pi}^{2\pi} \frac{r \cos^2 \phi d\phi}{1 + r^2 \cos ^2 \phi} \right) \notag \\
    & = \frac{r}{\pi} \int_{-\infty}^{0} \frac{du}{(1 + u^2)(1 + u^2 + r^2)} \tag{$u \coloneqq \tan \phi$}  \\
    & = \frac{1}{\pi r} \int_{-\infty}^{0} \left(\frac{1}{1 + u^2} - \frac{1}{1 + u^2 + r^2} \right) du \notag \\
    & = \frac{1}{\pi r} \left[ \mathrm{Arctan}\: u - \frac{1}{\sqrt{1+r^2}}\mathrm{Arctan}\left(\frac{u}{\sqrt{1+r^2}}\right) \right]_{-\infty}^{0} \notag \\
    & = \frac{1}{2r} \left(1 - \frac{1}{\sqrt{1+r^2}} \right). \label{sm1_g2}
\end{align}
In the derivation of Eq. \eqref{sm1_g2}, we use the relationships $\cos^2 \phi = \frac{1}{1+u^2}$ and $d\phi = \frac{du}{1+u^2}$.  
According to Eqs. \eqref{sm1_g1} and \eqref{sm1_g2}, we see that averaged equations (i.e., Eq. \eqref{average_2}) are calculated as 
\begin{subequations}
    \begin{align}
        \dot r &= -\frac{\eps \alpha}{2} r + \frac{\eps}{2\pi r} \log (1+r^2) \label{amp_sm_1},\\
        r \dot \theta &= -\frac{\eps}{2r}\left( 1 - \frac{1}{\sqrt{1+r^2}} \right) \label{ph_sm_1}.
    \end{align}
\end{subequations}

\begin{figure}
    \centering
    \includegraphics[width=.7\linewidth]{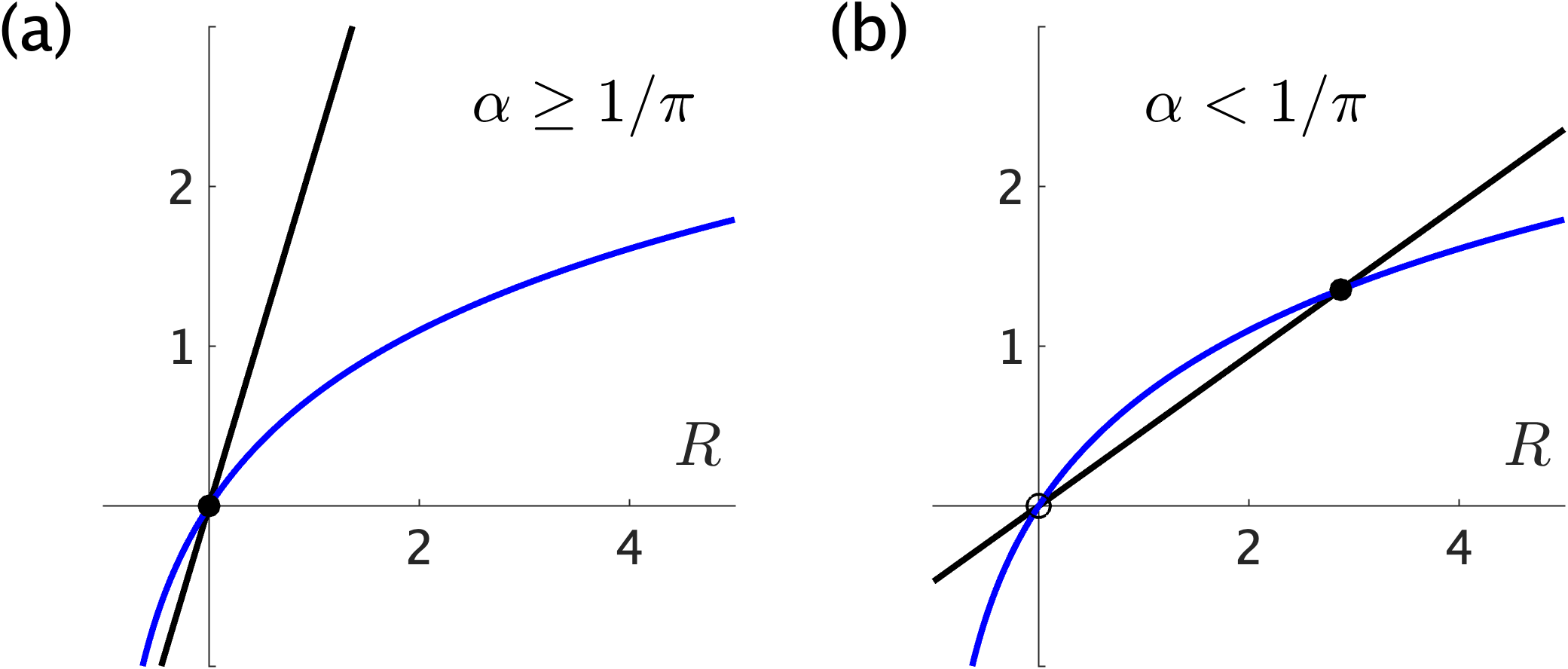}
    \caption{This figure shows the solutions of the transcendental equation \eqref{sm_1_fix_R}. The black and blue lines represent $\pi \alpha R$ and $\log(1+R)$, respectively. Both the black dot and black circle show the solutions of Eq. \eqref{sm_1_fix_R}. Note that the black dot corresponds to the stable fixed point of Eq. \eqref{amp_sm_1}, while the black circle corresponds to the unstable fixed point. ({\bf a}) If $\alpha$ equals to or is larger than $1/\pi$, which is the value when the black line is tangent to the blue curve, Eq. \eqref{sm_1_fix_R} has the unique solution $R=0$. We set $\alpha = 0.7$ to depict this panel. ({\bf b}) If $\alpha < 1/\pi$, Eq. \eqref{sm_1_fix_R} has another solution other than $R=0$. We set $\alpha = 0.15$ to depict this panel. }
    \label{fig_sm_1_R}
\end{figure}

We consider the dynamics of Eq. \eqref{amp_sm_1}. Obviously, the fixed point of Eq. \eqref{amp_sm_1} satisfies the following transcendental equation: 
\begin{equation}
    \label{sm_1_fix_R}
    \pi \alpha R = \log(1+R), 
\end{equation}
where $R \coloneqq r^2 \geq 0$. Since $\log(1+R)$ is a concave function of $R$, we see that Eq. \eqref{sm_1_fix_R} has the unique solution ($R=0$) if $\alpha \geq 1/\pi$ and two solutions if $\alpha < 1/\pi$ (Fig. \ref{fig_sm_1_R}). 

\begin{figure}
    \centering
    \includegraphics[width=.9\linewidth]{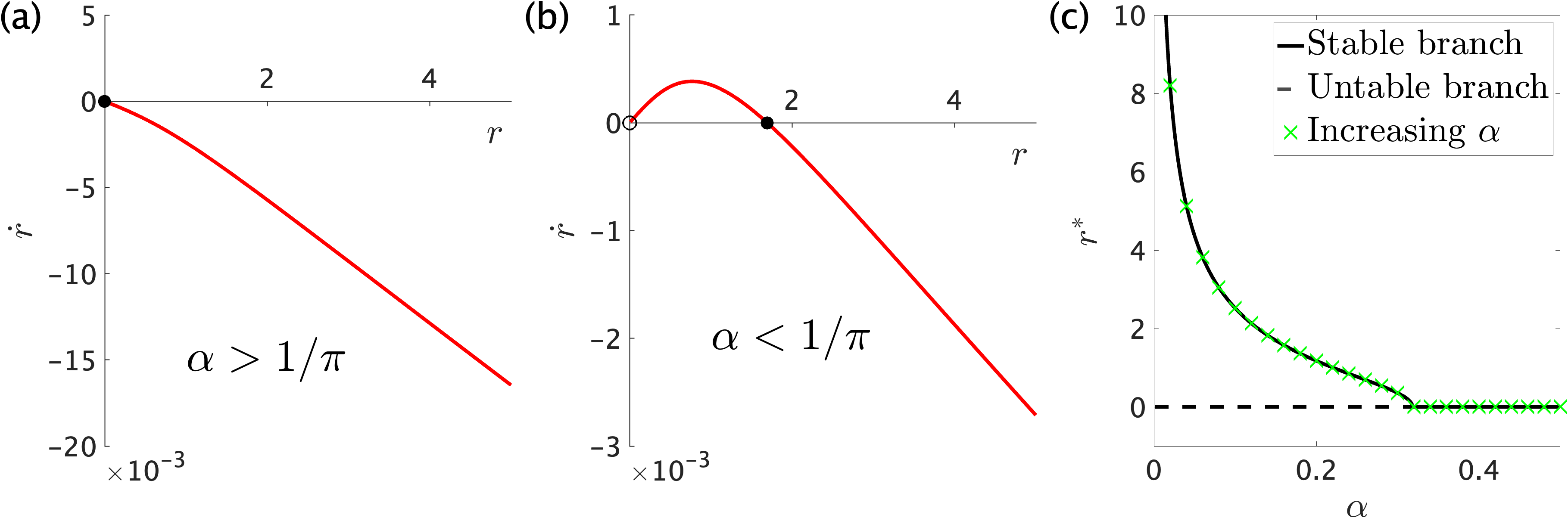}
    \label{single_flow_bif_sm1}
    \caption{({\bf a, b}) The typical flows of Eq. \eqref{amp_sm_1}, in which we plot $\dot r$ as a function of $r$. The black dot and black circle in each panel represent the stable and unstable fixed points, respectively. As for the parameter values, we set $\eps = 0.01$, $\alpha = 0.7$ in panel ({\bf a}), and $\alpha = 0.15$ in panel ({\bf b}). ({\bf c}) The bifurcation diagram for $r$ obtained by both the averaging approximation and the numerical simulation of Eq. \eqref{motion_eq_4'} where $g$ is given by Eq. \eqref{sm_1}. The solid and dashed lines represent the analytically obtained stable fixed point (the positive solution of Eq. \eqref{sm_1_fix_R} if $\alpha < 1/\pi$ and $r^*=0$ if $\alpha \geq 1/\pi$) and the unstable fixed point ($r^*=0$ if $\alpha < \pi$), respectively. The green cross marks show the equilibrium solutions obtained by numerical simulation of Eq. \eqref{motion_eq_4'}. We first use $\alpha = 0.02$ and then increase $\alpha$ by $0.02$ until $\alpha$ reaches to $0.5$. The initial condition for the first simulation is $x(0) = 2.0, \dot x(0) = 0$. For the following simulations, we use the equilibrium state of the previous simulation as the initial condition. We fix $\eps = 0.01$ for all the simulations in panel ({\bf c}). }
\end{figure}

Thus, as there exists a one-to-one relationship between $r \geq 0$ and $R$, we find the following: 
\begin{itemize}
    \item If $\alpha > 1/\pi$, Eq. \eqref{amp_sm_1} has a stable fixed point $r=0$. 
    \item The transcritical bifurcation occurs at $\alpha = 1/\pi$. 
    \item If $\alpha < 1/\pi$, Eq. \eqref{amp_sm_1} has an unstable fixed point ($r=0$) and a stable fixed point, whose value is the positive solution of Eq. \eqref{sm_1_fix_R}.
\end{itemize}
Figures \ref{single_flow_bif_sm1} ({\bf a}) and ({\bf b}) show the typical flows of Eq. \eqref{amp_sm_1} before and after the bifurcation point. We see that, as $\alpha$ decreases, the trivial fixed point $r=0$ changes to unstable and non-trivial fixed point emerges due to the transcritical bifurcation. 
The bifurcation diagram for $r$ is shown in Fig. \ref{single_flow_bif_sm1} ({\bf c}). The green cross marks show the numerically obtained equilibrium states of Eq. \eqref{motion_eq_4'} when we increase $\alpha$, which agree with the analytically obtained bifurcation diagram (black lines). 
However, in contrast to Model (i), (ii), and (iii) in the main article, this model is inappropriate to simulate the dynamics of metronome because it cannot simulate the bistability of the resting state and the oscillatory state. 

We consider that this problem arises because the escapement mechanism expressed as Eq. \eqref{sm_1} is not sufficiently weak near the resting state. In other words, Eq. \eqref{sm_1} implies that $g(x,y) \simeq x$ in the vicinity of $x=0$. These are the reasons why we use in the main article the rational function with numerator of degree 3 and denominator of degree 4 such that $g(x,y) \simeq 0$ holds in the vicinity of $x=0$ by a first-order approximation.


\section{Calculation of $\bar g_{1,2}(r)$ in the case where $g$ is given as Eq. \eqref{sm_2}}
\label{sm_2_app}
By substituting Eq. \eqref{change_variable2} into Eq. \eqref{sm_2}, we obtain
\begin{align}
    \label{sm_2_rt}
    g(r \cos \phi, -r \sin \phi) = 
    \begin{cases}
    \displaystyle\frac{r^3 \cos^3 \phi}{1 + r^4 \cos ^4 \phi} & 
    \begin{aligned}
        & {\rm if} \quad (2n + \textstyle\frac{1}{2}) \pi < \phi < (2n + 1) \pi \\
        & \quad {\rm or} \quad (2n - \textstyle\frac{1}{2}) \pi < \phi < 2n \pi, 
    \end{aligned} 
    \\
    \\
    0 & {\rm otherwise},
    \end{cases}
\end{align}
with arbitrary integer $n$.
Then, it follows from Eqs. \eqref{g12_def} and \eqref{sm_2_rt} that 
\begin{align}
    \bar g_1(r) &= \frac{1}{2 \pi} \left(\int_{\frac{\pi}{2}}^{\pi} \frac{r^3 \cos^3 \phi \sin \phi d\phi }{1 + r^4 \cos ^4 \phi} + \int_{\frac{3}{2}\pi}^{2\pi} \frac{r^3 \cos^3 \phi \sin \phi d\phi}{1 + r^4 \cos ^4 \phi} \right) \notag \\
    &= \frac{-1}{8 \pi r} \left\{ \left[ \log(1+r^4 \cos ^4 \phi) \right]_{\frac{\pi}{2}}^{\pi}  + \left[ \log(1+r^4 \cos ^4 \phi) \right]_{\frac{3}{2}\pi}^{2 \pi} \right\} \notag \\
    &= \frac{- \log(1+r^4)}{4 \pi r}, \label{sm2_g1}
\end{align}
and
\begin{align}
    \bar g_2(r) & = \frac{1}{2 \pi} \left(\int_{\frac{\pi}{2}}^{\pi} \frac{r^3 \cos^4 \phi d\phi }{1 + r^4 \cos ^4 \phi} + \int_{\frac{3}{2}\pi}^{2\pi} \frac{r^3 \cos^4 \phi d\phi}{1 + r^4 \cos^4 \phi} \right) \notag \\
    & = \frac{r^3}{\pi} \int_{-\infty}^{0} \frac{du}{(1 + u^2)\left[(1 + u^2)^2 + r^4\right]} \tag{$u \coloneqq \tan \phi$} \\
    & = \frac{1}{\pi r} \int_{-\infty}^{0} \left(\frac{1}{1 + u^2} - \frac{1+u^2}{(1 + u^2)^2 + r^4} \right) du \notag \\
    & = \frac{1}{\pi r} \left[ \mathrm{Arctan}\: u \right]_{-\infty}^{0} 
    - \frac{1}{\pi r}\int_{-\infty}^{0} \left( \frac{-\frac{\sqrt{\sqrt{1+r^4} - 1}}{2\sqrt{2(1+r^4)}}u + \frac{1}{2\sqrt{1 + r^4}}}{u^2 + \sqrt{2(\sqrt{1+r^4}-1)}u + \sqrt{1+r^4}} \right. 
    + \left. \frac{\frac{\sqrt{\sqrt{1+r^4} - 1}}{2\sqrt{2(1+r^4)}}u + \frac{1}{2\sqrt{1 + r^4}}}{u^2 - \sqrt{2(\sqrt{1+r^4}-1)}u + \sqrt{1+r^4}} \right) du \notag \\
    & 
    \begin{multlined}
        = \frac{1}{2r} - \frac{1}{\pi r} \left[ - \frac{\sqrt{\sqrt{1+r^4} - 1}}{4\sqrt{2(1+r^4)}} \log\left(u^2 + \sqrt{2(\sqrt{1+r^4}-1)}u + \sqrt{1+r^4}\right) \right. \notag \\
        + \frac{\sqrt{2 (\sqrt{1+r^4} + 1)}}{4\sqrt{1+r^4}} \mathrm{Arctan}\: \left( \sqrt{\frac{2}{1+\sqrt{1+r^4}}}\left(u + \sqrt{\frac{\sqrt{1+r^4} - 1}{2}}\right) \right)\notag \\
        + \frac{\sqrt{\sqrt{1+r^4} - 1}}{4\sqrt{2(1+r^4)}} \log\left(u^2 - \sqrt{2(\sqrt{1+r^4}-1)}u + \sqrt{1+r^4}\right) \notag \\
        + \left. \frac{\sqrt{2 (\sqrt{1+r^4} + 1)}}{4\sqrt{1+r^4}} \mathrm{Arctan}\: \left( \sqrt{\frac{2}{1+\sqrt{1+r^4}}}\left(u - \sqrt{\frac{\sqrt{1+r^4} - 1}{2}}\right) \right)\right]_{-\infty}^{0} \notag 
    \end{multlined}
    \\
    & = \frac{1}{2r} - \frac{\sqrt{2 (\sqrt{1+r^4} + 1)}}{4r\sqrt{1+r^4}} + \frac{\sqrt{\sqrt{1+r^4} - 1}}{4\sqrt{2(1+r^4)}} \lim_{u \to -\infty}\log \left(\frac{u^2 - \sqrt{2(\sqrt{1+r^4}-1)}u + \sqrt{1+r^4}}{u^2 + \sqrt{2(\sqrt{1+r^4}-1)}u + \sqrt{1+r^4}} \right) \notag \\
    & = \frac{1}{2r} \left( 1 - \sqrt{\frac{1+\sqrt{1+r^4}}{2(1 + r^4)}} \right).
    \label{sm2_g2}
\end{align}
According to Eqs. \eqref{sm2_g1} and \eqref{sm2_g2}, we see that Eq. \eqref{average_2} are calculated as Eqs. \eqref{amp_sm_2} and \eqref{ph_sm_2}.

\section{Calculation of $\bar g_{1,2}(r)$ in the case where $g$ is given as Eq. \eqref{poly}}
\label{poly_app}
By substituting Eq. \eqref{change_variable2} into Eq. \eqref{poly}, we obtain
\begin{align}
    \label{poly_rt}
    g(r \cos \phi, -r \sin \phi) = 
    \begin{cases}
    a r^3 \cos^3 \phi - b r^5 \cos ^5 \phi & 
    \begin{aligned}
        & {\rm if} \quad (2n + \textstyle\frac{1}{2}) \pi < \phi < (2n + 1) \pi \\
        & \quad {\rm or} \quad (2n - \textstyle\frac{1}{2}) \pi < \phi < 2n \pi, 
    \end{aligned} 
    \\
    \\
    0 & {\rm otherwise},
    \end{cases}
\end{align}
with arbitrary integer $n$.
Then, it follows from Eqs. \eqref{g12_def} and \eqref{poly_rt} that 
\begin{align}
    \bar g_1(r) &= \frac{1}{2 \pi} \left[ \int_{\frac{\pi}{2}}^{\pi} (a r^3 \cos^3 \phi - b r^5 \cos ^5 \phi) \sin \phi d\phi \right. \notag \\
    & \left.\qquad + \int_{\frac{3}{2}\pi}^{2\pi} (a r^3 \cos^3 \phi - b r^5 \cos ^5 \phi) \sin \phi d\phi \right] \notag \\
    &= \frac{-1}{\pi} \left[ \frac{a r^3 \cos^4 \phi}{4} - \frac{b r^5 \cos ^6 \phi}{6} \right]_{\frac{\pi}{2}}^{\pi} \notag \\
    &= \frac{- (3a r^3 - 2b r^5)}{12 \pi}, \label{poly_g1}
\end{align}
and
\begin{align}
    \bar g_2(r) &= \frac{1}{2 \pi} \left[ \int_{\frac{\pi}{2}}^{\pi} (a r^3 \cos^4 \phi - b r^5 \cos ^6 \phi) d\phi \right. \notag \\
    & \left.\qquad + \int_{\frac{3}{2}\pi}^{2\pi} (a r^3 \cos^4 \phi - b r^5 \cos ^6 \phi) d\phi \right] \notag \\
    &=\frac{1}{\pi}\left[ a r^3 \left( \frac{\sin 4\phi}{32} + \frac{\sin 2\phi}{4} + \frac{3 \phi}{8} \right) \right. \notag \\
    & \left. - b r^5 \left( \frac{\sin 6\phi}{192} + \frac{3 \sin 4\phi}{64} + \frac{15 \sin 2 \phi}{64} + \frac{5 \phi}{16} \right) \right]_{\frac{\pi}{2}}^{\pi} \notag \\
    &= \frac{6 a r^3 - 5 b r^5}{32}. \label{poly_g2}
\end{align}
According to Eqs. \eqref{poly_g1} and \eqref{poly_g2}, we see that Eq. \eqref{average_2} are calculated as Eqs. \eqref{amp_poly} and \eqref{ph_poly}.

\section{Derivation of Eq. \eqref{polar_eqs}}
We introduce the complex variables $A_i(t)$ that satisfy
\begin{subequations}
    \label{change_variable_two}
    \begin{align}
        x_i(t) &= \frac{1}{2} (A_i(t) e^{it} + c.c. ), \\
        y_i(t) &= \frac{1}{2} (iA_i(t) e^{it} + c.c. ). 
    \end{align}
\end{subequations}
Noting that $A_i = (x_i - iy_i)e^{-it}$, we transform Eq. \eqref{motion_eqs_5} into
\begin{equation}
    \label{amp_eq_two}
    \dot A_i = i e^{-it} \eps \left[ \mu(x_1 + x_2) + \beta y_i - g(x_i,y_i) \right]. 
\end{equation}
Equations \eqref{amp_eq_two} are derived in the same way as the one-oscillator system, which is described in Sec. \ref{derive_me} in Supplementary Information. 

We express $A_i(t)$ in a polar coordinate as below: 
\begin{equation}
    \label{polar_two}
    A_i(t) = r_i(t) e^{i \theta_i(t)},
\end{equation}
where $r_i(t), \theta_i(t)$ are real variables with $r_i(t) \geq 0$. Then, Eq. \eqref{change_variable2_two} follows from Eqs. \eqref{change_variable_two} and \eqref{polar_two}. 
By substituting Eqs. \eqref{polar_two} and \eqref{change_variable2_two} into Eq. \eqref{amp_eq_two}, we have
\begin{align}
    \dot r_i + i r_i \dot \theta_i &= i e^{-i \phi_i} \eps 
    \left[ \mu(r_1 \cos \phi_1 + r_2 \cos \phi_2) - \beta r_i \sin \phi_i  - g(r_i \cos \phi_i, -r_i \sin \phi_i) \right],
\end{align}
from which we obtain Eq. \eqref{polar_eqs}.

\section{Stability analysis of two-oscillator system}
\label{sec_stab_analy_app}
We perform the linear stability analysis for the fixed points $(r_1, r_2, \psi) = (r^*, r^*, 0)$ and $(r^*, r^*, \pi)$ under the condition \eqref{cond_OD}. The Jacobian matrix $J$ at these fixed points can be written in a unified manner as below: 
\begin{equation}
    J = 
    \begin{pmatrix}
        j_1 & 0 & - j_2 \\
        0 & j_1 & j_2 \\
        j_3 & -j_3 & 0
    \end{pmatrix}
    ,
\end{equation}
where
\begin{align}
    j_1 &\coloneqq \frac{\eps}{12 \pi} \left( - 6 \pi \beta + 9a r^{*2} -10b r^{*4} \right), \label{j1_two}\\
    j_2 &\coloneqq \frac{\eps \mu r^*s}{2}, \\
    j_3 &\coloneqq \frac{\eps}{8} \left( 3 a r^* - 5 b r^{*3} + \frac{8 \mu s}{r^*} \right), \label{j3_two}
\end{align}
and
\begin{equation}
    s \coloneqq
    \begin{cases}
        1 \quad {\rm if} \quad \psi = 0,  \\
        -1 \quad {\rm if} \quad \psi = \pi. 
    \end{cases}
\end{equation}
The eigenvalues of $J$ are 
\begin{equation}
    j_1, \quad \frac{j_1 \pm \sqrt{j_1^2 - 8 j_2 j_3}}{2}, 
\end{equation}
which implies that each of the fixed points \eqref{fixed_pt_two_in} and \eqref{fixed_pt_two_anti} is asymptotically stable if and only if 
\begin{equation}
    j_1 < 0 \quad {\rm and} \quad j_2 j_3 > 0. 
\end{equation}
Below, we examine the sign of $j_1, j_2$, and $j_3$. 

Obviously, $j_2>0$ if $\psi = 0$ and $j_2<0$ if $\psi = \pi$. By substituting Eq. \eqref{r*_two} into Eqs. \eqref{j1_two} and \eqref{j3_two}, we get
\begin{align}
    \label{j1_simp}
    j_1 &= - \frac{\eps}{8\pi b}\left( 3 a^2 - 16\pi b \beta + a \sqrt{9a^2 - 48\pi b \beta} \right), \\
    j_3 &= - \frac{\eps(27a^2 -120\pi b \beta + 9a\sqrt{9a^2 - 48\pi b \beta} - 64b\mu s)}{32\sqrt{(3a+\sqrt{9a^2 - 48\pi b \beta})b}}.\label{j3_simp}
\end{align}
It thus follows from inequality \eqref{cond_OD} and Eqs. \eqref{j1_simp} and \eqref{j3_simp} that 
\begin{equation}
    j_1 < 0, 
\end{equation}
and
\begin{equation}
    j_3 < 0 \quad {\rm if} \quad \psi = \pi.   
\end{equation}
If $\psi = 0$ (i.e., if $s=1$), then $j_3 > 0$ is true when inequality \eqref{j3_posi} holds. 

Based on the above discussion, we conclude that the fixed point $(r^*, r^*, 0)$ is always asymptotically stable and that the fixed point $(r^*, r^*, \pi)$ is asymptotically stable under the condition \eqref{j3_posi}. 
\newpage
\section{Supplementary figures}
We present several figures that support the main article. 
\begin{figure}[h]
    \centering
    \includegraphics[width=.6\linewidth]{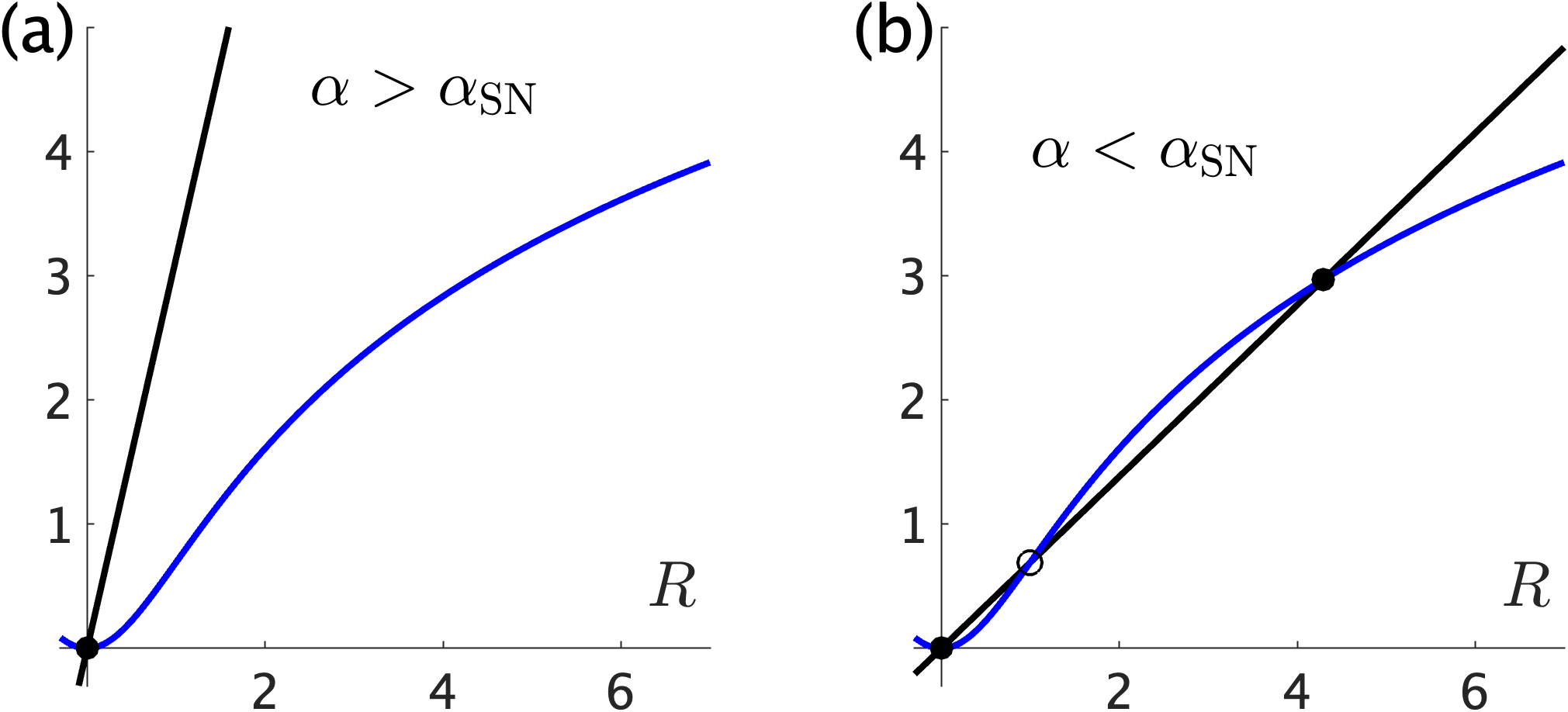}
    \caption{This figure shows the solutions of the transcendental equation \eqref{sm_2_fix_R}. The black and blue lines represent $2 \pi \alpha R$ and $\log(1+R^2)$, respectively. Both the black dot and black circle show the solutions of Eq. \eqref{sm_2_fix_R}. Note that the black dot corresponds to the stable fixed point of Eq. \eqref{amp_sm_2}, while the black circle corresponds to the unstable fixed point. ({\bf a}) If $\alpha$ is larger than $\alpha_{\rm SN}$, which is the value when the black line is tangent to the blue curve, Eq. \eqref{sm_2_fix_R} has the unique solution $R=0$. We set $\alpha = 0.5$ to depict this panel. ({\bf b}) If $\alpha < \alpha_{\rm SN}$, Eq. \eqref{sm_2_fix_R} has two solutions in addition to $R=0$. We set $\alpha = 0.11$ to depict this panel. }
    \label{fig_sm_2_R}
\end{figure}
\begin{figure}[h]
    \centering
    \includegraphics[width=.5\linewidth]{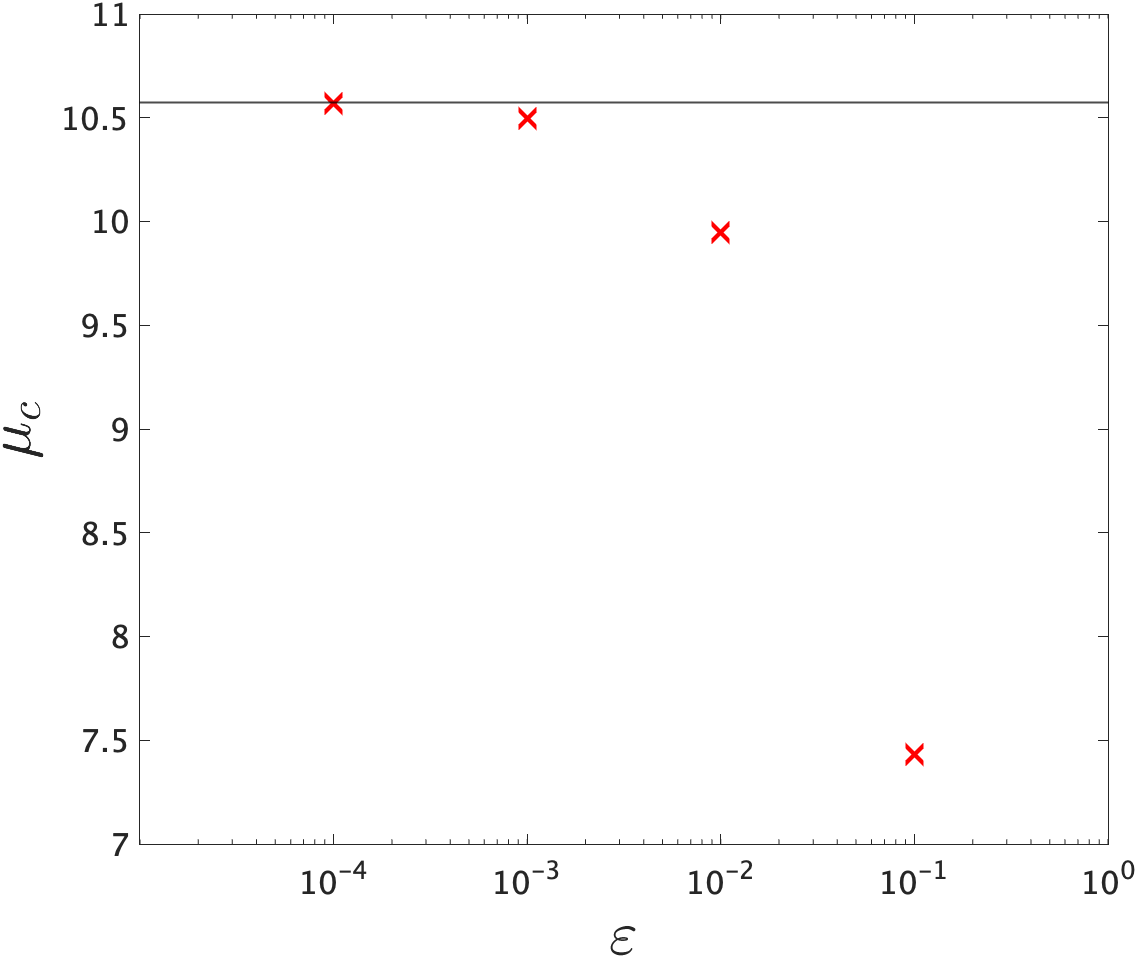}
    \caption{The relashionship between $\eps$ and $\mu_{\rm c}$, the value of $\mu$ at which the stability of the in-phase synchronization switches. The red cross marks show numerically obtained $\mu_{\rm c}$, derived by numerical integration of Eq. \eqref{motion_eqs_id_num} and bisection method. The black line shows $\mu_{\rm c}$ in inequality \eqref{j3_posi}, which is obtained by averaging approximation. Note that the red cross marks approach the black line as we decrease $\eps$. We fix $a=4, b=1, \beta = 0.3$ in this figure. }
    \label{asympt_graph}
\end{figure}
\makeatletter\@input{sub.tex}\makeatother